\documentclass[12pt]{article}

\usepackage[margin=0.8in]{geometry}
\usepackage{graphicx}
\usepackage{float} 
\usepackage{caption}
\usepackage{url}
\usepackage{mathtools} 
\usepackage{amsmath} 
\usepackage{amssymb} 
\usepackage{colortbl} 
\usepackage{multirow} 
\usepackage{array} 
\usepackage{cite, fancyhdr}
\usepackage{subcaption}
\usepackage{tikz}
\usetikzlibrary{patterns,angles,quotes,arrows}
\usepackage{xr, setspace, cuted}
\usepackage[inline]{enumitem}

\DeclareMathOperator{\Image}{Im}
\DeclareMathOperator{\rank}{rank}

\DeclareMathOperator{\relint}{relint}
\DeclareMathOperator{\interior}{int}
\DeclareMathOperator{\Span}{span}
\DeclareMathOperator{\co}{co}
\DeclareMathOperator{\proj}{proj}
\usepackage{amsthm}
\newtheoremstyle{mystyle}
  {}
  {}
  {}
  {}
  {\bfseries}
  {.}
  { }
  {}

\theoremstyle{mystyle}
\newtheorem{thm}{Theorem}
\newtheorem{prob}{Problem}
\newtheorem{defn}{Definition}
\newtheorem{prop}{Proposition}
\newtheorem{lem}{Lemma}
\newtheorem*{pf}{Proof}

\newtheorem{cor}{Corollary}

\begin{document}

\setlength{\textfloatsep}{0pt}
\setlength{\textfloatsep}{15pt plus 2pt minus 4pt}
\setlength{\textfloatsep}{8pt plus 2pt minus 4pt}
\setlength{\textfloatsep}{8pt plus 1pt minus 2pt}
\setlength{\dbltextfloatsep}{3pt}
\setlength{\intextsep}{5pt}
\setlength{\abovecaptionskip}{3pt plus 3pt minus 3pt}
\setlength{\belowcaptionskip}{3pt plus 3pt minus 3pt}
\setlength{\parskip}{4pt}

\setlength{\abovedisplayskip}{3pt}
\setlength{\belowdisplayskip}{3pt}
\setlength\abovedisplayshortskip{3pt}
\setlength\belowdisplayshortskip{3pt}

\setlist{nosep}

\renewcommand{\paragraph}[1]{\noindent {\bf #1}}

\title{Resilience of Linear Systems to Partial Loss of Control Authority}

\author{Jean-Baptiste~Bouvier and Melkior Ornik
\thanks{Jean-Baptiste Bouvier and Melkior Ornik are with the Department of Aerospace Engineering and the Coordinated Science Laboratory, University of Illinois at Urbana-Champaign, Urbana, IL 61801, USA.\hfill \break
e-mail: bouvier3@illinois.edu \& mornik@illinois.edu}}

\date{}

\maketitle

\begin{abstract} 
    After a loss of control authority over thrusters of the Nauka module, the International Space Station lost attitude control for 45 minutes with potentially disastrous consequences. Motivated by this scenario, we investigate the continued capability of control systems to perform their task despite partial loss of authority over their actuators. We say that a system is resilient to such a malfunction if for any undesirable inputs and any target state there exists an admissible control driving the state to the target. Building on controllability conditions and differential games theory, we establish a necessary and sufficient condition for the resilience of linear systems. As their task might be time-constrained, ensuring completion alone is not sufficient. We also want to estimate how much slower the malfunctioning system is compared to its nominal performance. Relying on Lyapunov theory we derive analytical bounds on the reach times of the nominal and malfunctioning systems in order to quantify their resilience. We illustrate our work on the ADMIRE fighter jet model and on a temperature control system.
\end{abstract}

\section{Introduction}

After the Nauka module docked to the International Space Station (ISS), a software failure caused a misfire of the module's thrusters, leading to a loss of attitude control of the whole station for 45 minutes \cite{ISS_thruster}. Eventually, other thrusters on the ISS were fired to counteract the uncontrolled and undesirable thrust until the Nauka module ran out of fuel. Motivated by such events, \cite{IFAC} introduced the notion of a \textit{partial loss of control authority over actuators} where some of the actuators of a system start producing uncontrolled and thus possibly undesirable inputs within their full range of actuation. To identify these faulty actuators, we assume sensors monitor each actuator in real time \cite{actuators_measures}. Our first objective is then one of \textit{resilient reachability}, i.e, verifying whether for all possible outputs of the malfunctioning actuators, the controlled ones can steer the system to its target \cite{IFAC}. Our second objective is to estimate the maximal time penalty caused by such a malfunction.

Classically, changing or unknown dynamics are studied through robust, adaptive, and fault-tolerant control theories. However, robust control needs the undesirable inputs to be significantly smaller than the controls \cite{weak_robust_control}. Since the loss of control authority over actuators may produce large undesirable inputs, robust control performs poorly \cite{TAC}. 
In turn, adaptive control tries to estimate unknown parameters before they have time to change significantly \cite{adaptive_control}, which may not be possible for uncontrolled inputs. Such a situation would typically prevent convergence of the estimators and lead to mediocre adaptive control performance \cite{weak_robust_control}.
As for fault-tolerant theory, actuator failure investigations are usually limited either to actuators ``locking in place" and producing constant inputs \cite{actuator_lock} or to actuators with reduced effectiveness \cite{loss_control_effectiveness, Fault_Tolerant_Review}. Since uncontrolled actuators can still produce a full range of inputs, loss of control authority over actuators is not covered by existing fault-tolerant theory \cite{Fault_Tolerant_Review}.

On the other hand, loss of control authority falls within the framework of differential games because the malfunctioning actuators can be modeled by adversaries as in \cite{Heymann_long, Schmitendorf_MaxMin}.
However, these works do not constitute appropriate starting points for a resilient reachability study due to the unbounded inputs of \cite{Heymann_long} and the complexity of the theory of \cite{Schmitendorf_MaxMin}.

Concerning our second objective, \textit{quantitative resilience} was introduced in \cite{SIAM_CT, Quantitative_Resilience} as the maximal ratio of the minimal reach times for the nominal and malfunctioning systems. However, the exact calculation of quantitative resilience for systems with driftless dynamics \cite{Maximax_Minimax_JOTA} does not extend to general linear systems since the minimal reach time in such systems does not have an analytical expression \cite{Athans}. 

The main contributions of this work are fourfold. 
Firstly, relying on the differential games theory of H\'ajek \cite{Hajek} and the controllability conditions of Brammer \cite{Brammer}, we establish simple necessary and sufficient conditions to verify the resilient stabilizability of linear systems, i.e., whether the origin is resiliently reachable from any initial state.
Secondly, we extend H\'ajek's duality theorem in order to study the resilient reachability of affine targets.
Thirdly, we use zonotopic underapproximations of reachable sets \cite{Zonotopes, Girard} to determine what states are guaranteed to be resiliently reachable.
Finally, we employ Lyapunov theory \cite{Kalman} to establish analytical bounds on the quantitative resilience of linear systems.

The remainder of this work is organized as follows. Section~\ref{sec: prelim} introduces the system dynamics and the problems of interest. Section~\ref{sec:background} provides background results.
Section~\ref{sec: stabilizability} establishes necessary and sufficient conditions for resilient stabilizability of linear systems. Section~\ref{sec: reachability} extends these conditions to affine targets and describes zonotopic underapproximations of the resiliently reachable set of linear systems.
Section~\ref{sec:quantitative} derives analytical bounds on the quantitative resilience of linear systems.
Section~\ref{sec:results} illustrates our theory on a fighter jet model and a temperature control system.

\textit{Notation:} We denote the integer interval from $a$ to $b$, inclusive, with $[\![a,b]\!]$. For a set $\Lambda \subseteq \mathbb{C}$, we say that $Re(\Lambda) \leq 0$ (resp. $Re(\Lambda) = 0$) if the real part of each $\lambda \in \Lambda$ verifies $Re(\lambda) \leq 0$ (resp. $Re(\lambda) = 0$).
The norm of a matrix $A$ is $\|A\| := \underset{x \neq 0}{\sup} \frac{\|Ax\|}{\|x\|} = \underset{\|x\| = 1}{\max} \|Ax\|$ and the set of its eigenvalues is $\lambda(A)$.
If $A$ is positive definite, denoted $A \succ 0$, then its extremal eigenvalues are $\lambda_{min}^A$ and $\lambda_{max}^A$, and $A$ generates a vector norm $\|x\|_A := \sqrt{x^\top Ax}$.
The controllability matrix of pair $(A,B)$ is $\mathcal{C}(A,B) = \big[B AB \hdots A^{n-1}B\big]$.
The zero matrix of size $n \times m$ is denoted by $0_{n, m}$, the identity matrix of size $n$ is $I_n$, the vector of ones is $\mathbf{1}_n$, and the vector of zeros except for a $1$ in position $i$ is $e_i$.
Set $\mathcal{Z}$ is symmetric if $-\mathcal{Z} = \mathcal{Z}$, its convex hull is denoted by $\co(\mathcal{Z})$, its interior by $\interior(\mathcal{Z})$, and its relative interior by $\relint(\mathcal{Z})$. 
The set of time functions taking value in $\mathcal{Z}$ is denoted $\mathcal{F}(\mathcal{Z}) := \big\{ f : f(t) \in \mathcal{Z}\ \text{for all}\ t \geq 0\big\}$.
The closed ball of dimension $b$, radius $r \geq 0$, and center $c$ is denoted $\mathbb{B}^b(c,r) := \big\{ x \in \mathbb{R}^b : \|x - c\| \leq r\big\}$.
The Minkowski addition of sets $\mathcal{X}$ and $\mathcal{Y}$ in $\mathbb{R}^n$ is $\mathcal{X} \oplus \mathcal{Y} := \big\{ x + y : x \in \mathcal{X}, y \in \mathcal{Y} \big\}$, and their Minkowski difference is $\mathcal{X} \ominus \mathcal{Y} := \big\{ z \in \mathbb{R}^n : \{z\} \oplus \mathcal{Y} \subseteq \mathcal{X} \big\}$. 
The projection map from $\mathbb{R}^n$ onto $\mathbb{R}^r$ with $r \leq n$ is denoted by $\proj_r(x_1, \hdots, x_n) := (x_1, \hdots, x_r) \in \mathbb{R}^r$.
The operator $\Span(\cdot)$ maps a set of vectors to their linear span.
The operator $\langle \cdot, \cdot \rangle$ denotes the standard scalar product in $\mathbb{R}^n$.

\section{Problem Statement}\label{sec: prelim}

We consider the linear time-invariant system
\begin{equation}\label{eq:initial ODE}
    \dot x(t) = Ax(t) + \bar{B} \bar{u}(t),\ x(0) = x_0 \in \mathbb{R}^n,\ \bar{u}(t) \in \bar{\mathcal{U}},
\end{equation}
with constant matrices $A \in \mathbb{R}^{n \times n}$ and $\bar{B} \in \mathbb{R}^{n \times (m+p)}$. The admissible controls are assumed to be in $\bar{\mathcal{U}} := [-1,1]^{m+p}$, in line with previous works \cite{ECC, Eaton, Kalman}. 

After a loss of control authority over $p$ of the $m+p$ actuators of system \eqref{eq:initial ODE}, the input signal $\bar{u}$ is split between the undesirable input signal $w \in \mathcal{F}(\mathcal{W})$, $\mathcal{W} := [-1, 1]^p$, and the controlled input signal $u \in \mathcal{F}(\mathcal{U})$, $\mathcal{U} := [-1, 1]^m$. Matrix $\bar{B}$ is accordingly split in $B \in \mathbb{R}^{n \times m}$ and $C \in \mathbb{R}^{n \times p}$ so that the dynamics become
\begin{equation}\label{eq:splitted ODE}
    \dot x(t) = Ax(t) + Bu(t) + Cw(t), \quad x(0) = x_0 \in \mathbb{R}^n.
\end{equation}

We want to study how the partial loss of control authority affects the \emph{stabilizability} and the \emph{controllability} of the nominal dynamics.

\begin{defn}
     System \eqref{eq:initial ODE} is \emph{stabilizable} (resp. \emph{controllable}) if there exists an admissible control signal $\bar{u} \in \mathcal{F}(\bar{\mathcal{U}})$ driving the state of system \eqref{eq:initial ODE} from any $x_0 \in \mathbb{R}^n$ to $0 \in \mathbb{R}^n$ (resp. to any $x_{tg} \in \mathbb{R}^n$).
\end{defn}

To adapt these two properties to system \eqref{eq:splitted ODE}, we first need the notion of \emph{resilient reachability} introduced in \cite{IFAC}.

\begin{defn}
    A target $x_{tg} \in \mathbb{R}^n$ is \emph{resiliently reachable} from $x_0 \in \mathbb{R}^n$ by system \eqref{eq:splitted ODE} if for all $w \in \mathcal{F}(\mathcal{W})$, there exists $T \geq 0$ and $u \in \mathcal{F}(\mathcal{U})$ such that $u(t)$ only depends on $w([0,t])$ and the solution to \eqref{eq:splitted ODE} exists, is unique, and $x(T) = x_{tg}$.
\end{defn}

Note that $u(t)$ is allowed to depend on $w(t)$ thanks to real time sensors on all actuators of the system, even on the malfunctioning ones.

\begin{defn}
    System \eqref{eq:splitted ODE} is \emph{resiliently stabilizable} (resp. \emph{resilient}) to the loss of the actuators corresponding to $C$ if $0 \in \mathbb{R}^n$ (resp. every $x_{tg} \in \mathbb{R}^n$) is resiliently reachable from any $x_0 \in \mathbb{R}^n$ by system \eqref{eq:splitted ODE}.
\end{defn}

We are now led to our first problem.

\begin{prob}\label{prob:resilience}
    Determine whether system \eqref{eq:splitted ODE} is resiliently stabilizable and/or resilient.
\end{prob}

Even if system \eqref{eq:splitted ODE} is not resilient, it might still be able to resiliently reach some targets, just not all of $\mathbb{R}^n$.

\begin{prob}\label{prob:reachable set}
    Determine the states $x_{tg} \in \mathbb{R}^n$ that are resiliently reachable from a given $x_0 \in \mathbb{R}^n$ by system \eqref{eq:splitted ODE}.
\end{prob}

For time-constrained missions, resilience is not sufficient. We also need to quantify how much slower the malfunctioning system is compared to the nominal one. To do so, we follow \cite{SIAM_CT} and introduce the \emph{nominal reach time}
\begin{equation}\label{eq: def T_N^*}
    T_N^*(x_0, x_{tg}) := \hspace{-1mm} \underset{\bar{u}\, \in\, \mathcal{F}(\bar{\mathcal{U}})}{\inf} \hspace{-1mm} \left\{   \begin{array}{c} T > 0 : x(T) = x_{tg}\\
    \text{in system}\ \eqref{eq:initial ODE} \end{array} \right\},
\end{equation}
the \emph{malfunctioning reach time}
\begin{equation}\label{eq: def T_M^*}
     T_M^*(x_0, x_{tg}) := \hspace{-2mm} \underset{w \in \mathcal{F}(\mathcal{W})}{\sup} \hspace{-0.8mm} \left\{ \hspace{-0.5mm} \underset{u \in \mathcal{F}(\mathcal{U})}{\inf} \hspace{-0.5mm} \left\{ \hspace{-1mm} \begin{array}{c}
         T > 0 : x(T) = x_{tg}\  \\
         \text{in system}\ \eqref{eq:splitted ODE}
    \end{array} \hspace{-1mm} \right\} \hspace{-0.7mm} \right\} \hspace{-0.5mm},
\end{equation}
and the \emph{quantitative resilience} 
\begin{equation}\label{eq:r_q}
    r_q(x_{tg}) := \underset{x_0\, \in\, \mathbb{R}^n}{\inf}\ \frac{T_N^*(x_0, x_{tg})}{T_M^*(x_0, x_{tg})}.
\end{equation}
If $x_0 = x_{tg}$, then $T_N^* = T_M^* = 0$ and we take the convention that their ratio is $1$. If $x_{tg}$ is reachable from $x_0$ by system \eqref{eq:initial ODE}, then Theorem~4.3 of \cite{Liberzon} states that the $\inf$ in \eqref{eq: def T_N^*} becomes $\min$ since $\bar{\mathcal{U}}$ is compact and convex. Similarly, $T_M^*$ in \eqref{eq: def T_M^*} is achieved by optimal signals $w^* \in \mathcal{F}(\mathcal{W})$ and $u^* \in \mathcal{F}(\mathcal{U})$ when system \eqref{eq:splitted ODE} is resilient.


The only way to calculate $u^*$ without any future knowledge of $w^*$ is to solve the intractable Isaac's main equation \cite{Borgest}, which is the differential games counterpart of the Hamilton-Jacobi-Bellman (HJB) equation. According to \cite{Isaacs_review}, Isaac's main equation is even more difficult to solve than the HJB equation, which usually results in intractable partial differential equations \cite{Liberzon}. Hence, \cite{Borgest} produces only suboptimal solutions, itself concluding that its practical contribution is minimal.

Instead of the setting of \cite{Borgest}, we choose \cite{Sakawa}, where $u^*$ and $w^*$ are unique, \textit{bang-bang} \cite{Rechtschaffen_equivalences}, and make a time-optimal transfer from $x_0$ to $x_{tg}$. The controller knows that $w^*$ will be chosen to make $T_M^*$ the longest. Thus, $u^*$ is chosen to react optimally to this worst undesirable input. Then, $w^*$ is chosen, and to make $T_M^*$ the longest, it is the same as the controller had predicted. Hence, from an outside perspective it appears as if the controller built $u^*$ knowing $w^*$ in advance, as reflected by \eqref{eq: def T_M^*}. Then, $T_M^*$ is time-optimal and can be meaningfully compared with $T_N^*$, leading to the following problem.

\begin{prob}\label{prob:r_q}
    Quantify the resilience of system \eqref{eq:splitted ODE}.
\end{prob}

We will now provide the background results upon which we build our theory.

\section{Background Results}\label{sec:background}

We first introduce H\'ajek's differential games approach \cite{Hajek} which relies on dynamics
\begin{equation}\label{eq: Hajek}
    \dot x(t) = Ax(t) + z(t), \quad x(0) = x_0 \in \mathbb{R}^n, \quad z(t) \in \mathcal{Z},
\end{equation}
where $\mathcal{Z} \subseteq \mathbb{R}^n$ is the Minkowski difference between the set of admissible control inputs $B\mathcal{U} := \big\{ Bu : u \in \mathcal{U}\big\}$ and the opposite of the set of undesirable inputs $C\mathcal{W} := \big\{ Cw : w \in \mathcal{W}\big\}$, i.e.,
\begin{align*}
    \mathcal{Z} := B\mathcal{U} \ominus (-C\mathcal{W}) = \big\{ z \in B\mathcal{U} : z - Cw \in B\mathcal{U}\ \text{for all}\ w \in \mathcal{W} \big\}.
\end{align*}

\begin{thm}[H\'ajek's duality theorem \cite{Hajek}]\label{thm:Hajek}
    The state of system \eqref{eq:splitted ODE} can be driven to $0 \in \mathbb{R}^n$ at time $T$ for all $w \in \mathcal{F}(\mathcal{W})$ by control signal $u \in \mathcal{F}(\mathcal{U})$ if and only if the state of system \eqref{eq: Hajek} can be driven to $0$ at time $T$ by a control signal $z \in \mathcal{F}(\mathcal{Z})$, and $Bu(\cdot) = z(\cdot) - Cw(\cdot)$.
\end{thm}

Informally, $\mathcal{Z}$ represents the control available after counteracting any undesirable input.
Since $\bar{\mathcal{U}}$ is symmetric, compact, and convex, sets $B\mathcal{U}$ and $C\mathcal{W}$ also have these properties by linearity. According to \cite{Pontryagin_difference}, $\mathcal{Z}$ is then also symmetric, compact, and convex.

Theorem~\ref{thm:Hajek} transforms the resilient stabilizability of system \eqref{eq:splitted ODE} into the stabilizability of system \eqref{eq: Hajek}.
Because inputs are bounded, Kalman's stabilizability condition \cite{Heymann_long} do not apply, instead we employ Corollary~3.6 of \cite{Brammer}.

\begin{thm}[Stabilizability condition \cite{Brammer}]\label{thm:Brammer stabilizability}
    If $\bar{\mathcal{U}} \cap \ker(\bar{B}) \neq \emptyset$ and $\interior(\co(\bar{\mathcal{U}})) \neq \emptyset$, then system \eqref{eq:initial ODE} is stabilizable if and only if $\rank\big( \mathcal{C}(A,\bar{B}) \big) = n$, $Re\big(\lambda(A)\big) \leq 0$, and there is no real eigenvector $v$ of $A^\top$ satisfying $v^\top \bar{B} \bar{u} \leq 0$ for all $\bar{u} \in \bar{\mathcal{U}}$.
\end{thm}

The first condition of Theorem~\ref{thm:Brammer stabilizability} ensures the existence of a control canceling $\bar{B}\bar{u}$ so that the state can be maintained at an equilibrium. 
The rank condition is Kalman's \cite{Brammer} and the last two conditions guarantee that the drift term $Ax$ does not prevent stabilization. If $\bar{\mathcal{U}} = \mathbb{R}^m$, Theorem~\ref{thm:Brammer stabilizability} reduces to the usual stabilizability condition.

To verify controllability we use Corollary 3.7 of \cite{Brammer}, which is very similar to Theorem~\ref{thm:Brammer stabilizability} except that the eigenvalues of $A$ must have a zero real part to avoid creating a drift preventing the reachability of affine targets.

\begin{thm}[Controllability condition \cite{Brammer}]\label{thm:Brammer controllability}
    If $\bar{\mathcal{U}} \cap \ker(\bar{B}) \neq \emptyset$ and $\interior(\co(\bar{U})) \neq \emptyset$, then system~\eqref{eq:initial ODE} is controllable if and only if $\rank\big( \mathcal{C}(A, \bar{B}) \big) = n$, $Re\big(\lambda(A)\big) = 0$, and there is no real eigenvector $v$ of $A^\top$ satisfying $v^\top \bar{B} \bar{u} \leq 0$ for all $\bar{u} \in \bar{\mathcal{U}}$.
\end{thm}

We now have all the background results to start solving Problem~\ref{prob:resilience} by investigating resilient stabilizability.

\section{Resilient Stabilizability}\label{sec: stabilizability}

In this section, we first establish a simple resilient stabilizability condition before deriving a more complex condition with a wider range of application.

\begin{prop}\label{prop: resilient stabilizability Z}
    If $\interior(\mathcal{Z}) \neq \emptyset$, then system \eqref{eq:splitted ODE} is resiliently stabilizable if and only if $Re\big(\lambda(A)\big) \leq 0$.
\end{prop} \vspace{-3mm}
\begin{pf}
    According to Theorem~\ref{thm:Hajek}, the resilient stabilizability of system \eqref{eq:splitted ODE} is equivalent to the stabilizability of system \eqref{eq: Hajek}. We apply Theorem~\ref{thm:Brammer stabilizability} and obtain that if $\mathcal{Z} \cap \ker(I) \neq \emptyset$ and $\interior(\co(\mathcal{Z})) \neq \emptyset$ in $\mathbb{R}^n$, then system \eqref{eq: Hajek} is stabilizable if and only if $\rank\big( \mathcal{C}(A,I) \big) = n$, $Re\big(\lambda(A)\big) \leq 0$, and there is no real eigenvector $v$ of $A^\top$ satisfying $v^\top I z \leq 0$ for all $z \in \mathcal{Z}$.
    
    Because $\ker(I) = \{0\}$, the first condition becomes $0 \in \mathcal{Z}$.
    Since $\mathcal{Z}$ is convex, the second condition becomes $\interior(\mathcal{Z}) \neq \emptyset$, which is equivalent to $0 \in \interior(\mathcal{Z})$ according to Lemma~\ref{lemma: int Z non-empty iff 0 in Z} of Appendix~\ref{apx:lemmas}. This second condition implies the first one, so we only keep $\interior(\mathcal{Z}) \neq \emptyset$.
    
    We now assume that $\interior(\mathcal{Z}) \neq \emptyset$ and we simplify the last three conditions.
    Since $\rank(I) = n$, the third condition is always true.
    Lemma~\ref{lemma: int Z non-empty iff 0 in Z} yields $0 \in \interior(\mathcal{Z})$. Thus, there exists $\varepsilon > 0$ such that $\mathbb{B}^n(0, \varepsilon) \subseteq \mathcal{Z}$. If $A^\top$ has no real eigenvector, the last condition is trivially true. Otherwise, for $v$ be a real eigenvector of $A^\top$. Let $z = \varepsilon \frac{v}{\|v\|}$, then $z \in \mathbb{B}^n(0, \varepsilon)$, so $z \in \mathcal{Z}$ and $v^\top I z = \varepsilon \|v\| > 0$. $\quad \blacksquare$
\end{pf}

Proposition~\ref{prop: resilient stabilizability Z} has a limited range of application because of its requirement $\interior(\mathcal{Z}) \neq \emptyset$ in $\mathbb{R}^n$, i.e., $\mathcal{Z}$ must be of dimension $n$. However, stabilizability does not require $B\mathcal{U}$ to be dimension $n$, so resilient stabilizability should not require that from $\mathcal{Z}$ either. We then want our condition to rely on the relative interior of $\mathcal{Z}$ instead of its interior.

\begin{defn}
     The \emph{relative interior} $\relint(\mathcal{S})$ of a set $\mathcal{S}$ is the interior of $\mathcal{S}$ considered as a subset of its affine hull.
\end{defn}

\begin{defn}
    The \emph{affine hull} of a set $\mathcal{S}$ is the largest subspace included in $\mathcal{S}$ with respect to inclusion.
\end{defn}

If we apply Theorem~\ref{thm:Brammer stabilizability} to system \eqref{eq: Hajek} as in Proposition~\ref{prop: resilient stabilizability Z}, then $\interior(\mathcal{Z}) \neq \emptyset$ will appear.
Instead, we first need to transport system \eqref{eq: Hajek} into a basis adapted to $\mathcal{Z}$. 
Let $r := \dim(\mathcal{Z}) \leq n$. If $\mathcal{Z} = \emptyset$, we take the convention that $r = -\infty$ and $Z := [\, ] \in \mathbb{R}^{n \times 0}$, the empty matrix with $\Image([\, ]) = \emptyset$. Otherwise, according to Lemma~\ref{lemma: relint Z non-empty iff 0 in Z} of Appendix~\ref{apx:lemmas}, we have $0 \in \mathcal{Z}$. Then, $\Span(\mathcal{Z})$ is a vector space from which we take a basis $\{z_1, \hdots, z_r\}$ in $\mathbb{R}^n$. We define the matrix $Z := \big(z_1, \hdots, z_r\big) \in \mathbb{R}^{n \times r}$ with the convention that $Z = 0 \in \mathbb{R}^{n \times 1}$ if $r = 0$. 
Then, $\Image(Z) = \Span(\mathcal{Z})$ and we can formulate a resilient stabilizability condition less restrictive than Proposition~\ref{prop: resilient stabilizability Z}.

\begin{prop}\label{prop: resilient stabilizability}
    If $\relint(\mathcal{Z}) \neq \emptyset$, then system \eqref{eq:splitted ODE} is resiliently stabilizable if and only if \break
    $Re\big(\lambda(A)\big) \leq 0$, $\rank\big( \mathcal{C}(A,Z) \big) = n$, and there is no real eigenvector $v$ of $A^\top$ satisfying $v^\top z \leq 0$ for all $z \in \mathcal{Z}$.
\end{prop}
\begin{pf}
    We apply Theorem~\ref{thm:Hajek} and work on system \eqref{eq: Hajek}. Since $z_1, \hdots, z_r$ are linearly independent, we complete this sequence into a basis of $\mathbb{R}^n$ with $V := (v_{r+1},\hdots,v_n)$ and obtain a transition matrix $T_z = (Z,V)$. We change basis in system \eqref{eq: Hajek} with $x = T_z^{-1} y$ so that $\dot x(t) = T_z^{-1} \dot y(t) = T_z^{-1}Ay(t) + T_z^{-1} z(t) = \hat{A}x(t) + s(t)$, with $\hat{A} = T_z^{-1} A T_z$ and $s(t) \in \mathcal{S} := T_z^{-1}\mathcal{Z} = \big\{ T_z^{-1}z : z \in \mathcal{Z} \big\}$. By definition, $z_i = T_z e_i$ and thus $\mathcal{S} \subseteq \Span(\{e_1, \hdots, e_r\})$ in $\mathbb{R}^n$. Let $s \in \mathcal{S}$. Then,
    \begin{equation*}
        s = \def\arraystretch{0.7}\begin{pmatrix} s_1 \\ \vdots \\ s_r \\ 0_{n-r, 1} \end{pmatrix} = \begin{pmatrix} I_r \\ 0_{n-r, r} \end{pmatrix} \begin{pmatrix} s_1 \\ \vdots \\ s_r \end{pmatrix} := \hat{B} \hat{s},
    \end{equation*}
    with $\hat{B} = T_z^{-1}Z \in \mathbb{R}^{n \times r}$ and $\hat{s} \in \mathbb{R}^r$, $\hat{s} \in \hat{\mathcal{S}} := \proj_r(\mathcal{S})$, the projection of $\mathcal{S}$ onto $\mathbb{R}^r$. Hence, the stabilizability of system \eqref{eq: Hajek} is equivalent to that of system
    \begin{equation}\label{eq:basis T ODE}
        \dot{\hat{x}}(t) = \hat{A} \hat{x}(t) + \hat{B} \hat{s}(t), \quad \hat{x}(0) = T_z^{-1}x_0, \quad \hat{s}(t) \in \hat{\mathcal{S}}. 
    \end{equation}
    Applying Theorem~\ref{thm:Brammer stabilizability} to system \eqref{eq:basis T ODE} leads to the following stabilizability conditions: $\hat{\mathcal{S}} \cap \ker(\hat{B}) \neq \emptyset$, $\interior(\co(\hat{\mathcal{S}})) \neq \emptyset$, $Re(\lambda(\hat{A})) \leq 0$, $\rank\big( \mathcal{C}(\hat{A}, \hat{B}) \big) = n$, and there is no real eigenvector $\hat{v}$ of $\hat{A}^\top$ satisfying $\hat{v}^\top \hat{B} \hat{s} \leq 0$ for all $\hat{s} \in \hat{\mathcal{S}}$. We now simplify these five conditions.
    
    \begin{enumerate}
        \item Since $\hat{B} = \left( \begin{smallmatrix} I_r \\ 0 \end{smallmatrix} \right)$, $\rank(\hat{B}) = r$, and hence $\ker(\hat{B}) = \{0\}$ in $\mathbb{R}^r$. Then, $\hat{\mathcal{S}} \cap \ker(\hat{B}) \neq \emptyset$ is equivalent to $0 \in \hat{\mathcal{S}} = \proj_r(T_z^{-1} \mathcal{Z})$. In turn, this is equivalent to the existence of $v \in \mathbb{R}^{n-r}$ such that $T_z \left( \begin{smallmatrix} 0 \\ v \end{smallmatrix} \right) \in \mathcal{Z}$, i.e., $Vv \in \mathcal{Z}$. By definition of $V$, $\Image(V) \cap \Span(\mathcal{Z}) = \{0\}$. Thus, $\hat{\mathcal{S}} \cap \ker(\hat{B}) \neq \emptyset$ is equivalent to $0 \in \mathcal{Z}$, i.e., $\relint(\mathcal{Z}) \neq \emptyset$ according to Lemma~\ref{lemma: relint Z non-empty iff 0 in Z} of Appendix~\ref{apx:lemmas}.
        
        \item By definition of $\mathcal{S}$, $\interior(\hat{\mathcal{S}}) \neq \emptyset$ in $\mathbb{R}^r$ is equivalent to $\relint(\mathcal{Z}) \neq \emptyset$ since $T_z$ is invertible. 
        
        \item Because $\hat{A} = T_z^{-1} A T_z$, $\lambda(A) = \lambda(\hat{A})$, and thus the third condition becomes $Re(\lambda(A)) \leq 0$.
        
        \item For $i \hspace{-0.5mm} \in \hspace{-0.5mm} [\![0,n \hspace{-0.6mm} - \hspace{-0.6mm} 1]\!]$, $T_z \hat{A}^i \hspace{-0.5mm} \hat{B} \hspace{-0.5mm} = \hspace{-0.5mm} T_z \hspace{-0.5mm} \big( T_z^{-1} \hspace{-0.5mm} A T_z\big)^i \hspace{-0.5mm} \hat{B} \hspace{-0.5mm} = \hspace{-0.5mm} A^i T_z \hat{B} \hspace{-0.5mm} = A^i Z$ because $T_z\hat{B} = Z$. Hence, $\Image \big( T_z \mathcal{C}(\hat{A}, \hat{B}) \big) = \Image \big(  \mathcal{C}(A, Z) \big)$. Then, the invertibility of $T_z$ leads to $\rank \big( \mathcal{C}(\hat{A}, \hat{B}) \big) = \rank \big( \mathcal{C}(A, Z) \big)$ \cite{matrix_computations}.
        
        \item Assume that $\hat{v}$ is a real eigenvector of $\hat{A}^\top$ associated to the eigenvalue $\hat{\lambda}$. 
        Then, $v := T_z^{-\top} \hat{v}$ is an eigenvector of $A^\top$ associated to the same eigenvalue $\hat{\lambda}$ \cite{matrix_computations}. For $\hat{s} \in \hat{\mathcal{S}}$, we have $\hat{B} \hat{s} \in \mathcal{S}$ by definition. Hence, if we define $z := T_z \hat{B} \hat{s}$, we have $z \in \mathcal{Z}$. Then, $\hat{v}^\top \hat{B} \hat{s} = v^\top T_z \hat{B} \hat{s} = v^\top z$. $\quad \blacksquare$
    \end{enumerate}
\end{pf}

To further expand the applicability of our resilient stabilizability condition, we now remove the requirement $\relint(\mathcal{Z}) \neq \emptyset$ from Proposition~\ref{prop: resilient stabilizability} and obtain a necessary and sufficient condition.

\begin{thm}[Resilient stabilizability condition]\label{thm: N&S resilient stabilizability}
    System \eqref{eq:splitted ODE} is resiliently stabilizable if and only if $\rank\big( \mathcal{C}(A, Z) \big) = n$, $Re\big(\lambda(A)\big) \leq 0$, and there is no real eigenvector $v$ of $A^\top$ satisfying $v^\top z \leq 0$ for all $z \in \mathcal{Z}$.
\end{thm}
\begin{pf}
    Let us define the three properties stated in Proposition~\ref{prop: resilient stabilizability} as $\mathcal{P}_1 := $``$\relint(\mathcal{Z}) \neq \emptyset$", $\mathcal{P}_2 := $``System \eqref{eq:splitted ODE} is resiliently stabilizable", and $\mathcal{P}_3 :=$``$\rank \hspace{-0.8mm}\big( \mathcal{C}(A, Z) \hspace{-0.3mm} \big) \hspace{-0.6mm} = n$, $Re\big(\lambda(A)\big) \leq 0$, and there is no real eigenvector $v$ of $A^\top$ satisfying $v^\top z \leq 0$ for all $z \in \mathcal{Z}$".
    Proposition~\ref{prop: resilient stabilizability} states that if $\mathcal{P}_1$ holds, then $\mathcal{P}_2$ is equivalent to $\mathcal{P}_3$. We will now show that when $\mathcal{P}_1$ is false, so are $\mathcal{P}_2$ and $\mathcal{P}_3$, which leads to $\mathcal{P}_2$ equivalent to $\mathcal{P}_3$ no matter the status of $\mathcal{P}_1$, which is exactly the statement of this theorem.
    
    Assume that $\mathcal{P}_1$ is false. Then, according to Lemmas~\ref{lemma: relint Z non-empty iff 0 in Z}, \ref{lemma: Z = empty}, and \ref{lemma: not stabilizable} of Appendix~\ref{apx:lemmas}, system \eqref{eq:splitted ODE} is not resiliently stabilizable, i.e., $\mathcal{P}_2$ is false. We took the convention that $Z = [\, ]$ with $\rank([\, ]) = -\infty$, so $\mathcal{P}_3$ is false too. $\quad \blacksquare$
\end{pf}

Note that the rank condition in Theorem~\ref{thm: N&S resilient stabilizability} concerns the pair $(A,Z)$ and not $(A,B)$ as one might have wanted. For the stabilizability of these pairs to be equivalent, we need $\mathcal{Z}$ and $B\mathcal{U}$ to have the same dimension.

\begin{cor}\label{cor: resilient stabilizability}
    If $\dim(\mathcal{Z}) = \rank(B)$, then system \eqref{eq:splitted ODE} is resiliently stabilizable if and only if $\rank\big( \mathcal{C}(A, B) \big) = n$, $Re\big(\lambda(A)\big) \leq 0$, and there is no real eigenvector $v$ of $A^\top$ satisfying $v^\top z \leq 0$ for all $z \in \mathcal{Z}$.
\end{cor}
\begin{pf}
    If $\mathcal{Z} = \emptyset$, then $\rank(B) = -\infty$, i.e., $B = []$, so \eqref{eq:splitted ODE} is not resiliently stabilizable and $\rank\big( \mathcal{C}(A, B) \big) \neq n$.

    Now assume that $\mathcal{Z} \neq \emptyset$. From Lemma~\ref{lemma: span Z = Im B} of Appendix~\ref{apx:lemmas} we get $\Image(B) = \Image(Z)$. Then, $\Image \big( \mathcal{C}(A, B) \big) = \Image \big( \mathcal{C}(A, Z) \big)$. In the proof of Proposition~\ref{prop: resilient stabilizability} we had $\Image \big( \mathcal{C}(A, Z) \big) = \Image \big( T \mathcal{C}(\hat{A}, \hat{B}) \big)$. Since $T$ is invertible, we obtain $\rank\big( \mathcal{C}(A, B) \big) = \rank\big( \mathcal{C}(\hat{A}, \hat{B}) \big)$, and we conclude with the rest of the proof of Proposition~\ref{prop: resilient stabilizability}. $\quad \blacksquare$
\end{pf}

Notice how the three conditions listed in Corollary~\ref{cor: resilient stabilizability} are similar to the stabilizability conditions from Theorem~\ref{thm:Brammer stabilizability}. We are then led to the following result.

\begin{cor}\label{cor: res stab = initial stab}
    If $\dim(\mathcal{Z}) = \rank(B)$, then system \eqref{eq:splitted ODE} is resiliently stabilizable if and only if system \eqref{eq:initial ODE} is stabilizable.
\end{cor}
\begin{pf}
    Let $v$ be a real eigenvector of $A^\top$. Assume first that there exists $z \in \mathcal{Z}$ such that $v^\top z > 0$. By construction of $B$, $\mathcal{U}$, and $\mathcal{Z}$, we have $\mathcal{Z} \subseteq B\mathcal{U} \subseteq \bar{B} \bar{\mathcal{U}}$. Hence, there exists $\bar{u} \in \bar{\mathcal{U}}$ such that $z = \bar{B}\bar{u}$ and $v^\top \bar{B} \bar{u} > 0$.

    On the other hand, assume that there exists $\bar{u} \in \bar{\mathcal{U}}$ such that $v^\top \bar{B} \bar{u} > 0$. According to Lemma~\ref{lemma: span Z = Im B}, $\Span(\mathcal{Z}) = \Image(\bar{B})$. Then, the convexity of $\mathcal{Z}$ yields the existence of $\alpha \in \mathbb{R}$ and $z \in \mathcal{Z}$ such that $\bar{B} \bar{u} = \alpha z$. Note that $\alpha \neq 0$ by definition of $\bar{u}$. If $\alpha > 0$, we have $v^\top z > 0$. Otherwise, $\alpha < 0$ but we use the symmetry of $\mathcal{Z}$ to obtain $-z \in \mathcal{Z}$ and $v^\top (-z) > 0$.
    
    Thus, the condition ``there is no real eigenvector $v$ of $A^\top$ satisfying $v^\top z \leq 0$ for all $z \in \mathcal{Z}$" is equivalent to ``there is no real eigenvector $v$ of $A^\top$ satisfying $v^\top \bar{B}\bar{u} \leq 0$ for all $\bar{u} \in \bar{\mathcal{U}}$" when $\dim(\mathcal{Z}) = \rank(B)$. According to Lemma~\ref{lemma: span Z = Im B} of Appendix~\ref{apx:lemmas}, $\Image(B) = \Image(\bar{B})$. Hence, $\rank\big( \mathcal{C}(A, B) \big) = \rank\big( \mathcal{C}(A, \bar{B}) \big)$.
    Then, applying Corollary~\ref{cor: resilient stabilizability} to system~\eqref{eq:splitted ODE} and Theorem~\ref{thm:Brammer stabilizability} to system~\eqref{eq:initial ODE} concludes the proof. $\quad \blacksquare$
\end{pf}

We have established several resilient stabilizability conditions, hence solving the first half of Problem~\ref{prob:resilience}. We will now tackle its second part concerning affine targets.

\section{Resilient Reachability}\label{sec: reachability}

In this section we extend H\'ajek's duality theorem \cite{Hajek} to affine targets and study the resilience of linear systems.

\begin{thm}[Extended duality theorem]\label{thm:Hajek non-zero}
    The state of system \eqref{eq:splitted ODE} can be driven to $x_{tg} \in \mathbb{R}^n$ at time $T$ for all $w \in \mathcal{F}(\mathcal{W})$ by control signal $u \in \mathcal{F}(\mathcal{U})$ if and only if the state of system \eqref{eq: Hajek} can be driven to $x_{tg}$ at time $T$ by a control signal $z \in \mathcal{F}(\mathcal{Z})$, and $Bu(\cdot) = z(\cdot) - Cw(\cdot)$.
\end{thm}
\begin{pf}
     Consider system \eqref{eq:splitted ODE} with a target state $x_{tg} \in \mathbb{R}^n$, $x_{tg} \neq 0$. Let $X(t) := \left( \begin{smallmatrix} x(t) - x_{tg} \\ A x_{tg} \end{smallmatrix} \right) \in \mathbb{R}^{2n}$. Then,
    \begin{equation}\label{eq:X ODE split}
    \begin{array}{c}
         \dot X(t) = A_2 X(t) + B_2 u(t) + C_2 w(t), \\
         X(0) = X_0 \in \mathbb{R}^{2n}, \quad u(t) \in \mathcal{U}, \quad w(t) \in \mathcal{W}, 
    \end{array}
    \end{equation}
    \begin{equation*}
        \text{with} \quad A_2 =  \begin{pmatrix} A & I_n \\ 0_{n, n} & 0_{n, n} \end{pmatrix}, \quad B_2 = \begin{pmatrix} B \\ 0_{n, m} \end{pmatrix}, \quad C_2 = \begin{pmatrix} C \\ 0_{n, p} \end{pmatrix}, \quad \text{and} \quad X_0 = \begin{pmatrix} x_0 - x_{tg} \\ Ax_{tg} \end{pmatrix}.
    \end{equation*}
     Let the target set be $\mathcal{G} = \big\{ \left( \begin{smallmatrix} 0 \\ a \end{smallmatrix} \right) \in \mathbb{R}^{2n} \big\} = \{0\}^n \times \mathbb{R}^n$. 
    
    Since $0 \in C_2\mathcal{W}$, we can apply H\'ajek's second duality theorem of \cite{Hajek} stating that $\mathcal{G}$ is resiliently reachable in time $T$ from $X_0$ by system \eqref{eq:X ODE split} if and only if $\mathcal{G}$ is reachable in time $T$ from $X_0$ by the following system
    \begin{equation}\label{eq:X Hajek ODE}
         \dot X(t) = A_2 X(t) + v_2(t), \quad X(0) = X_0, 
    \end{equation}
    \begin{equation*}
        v_2(t) \in \mathcal{V}_2 := B_2 \mathcal{U} \cap \big[ (B_2 \mathcal{U} \oplus \mathcal{G}_{A_2}) \ominus (-C_2 \mathcal{W}) \big] \subseteq \mathbb{R}^{2n},
    \end{equation*}
    where $\mathcal{G}_{A_2}$ is the largest subspace of $\mathcal{G}$ invariant by $A_2$. Take $g = \left( \begin{smallmatrix} 0 \\ a \end{smallmatrix} \right) \in \mathcal{G}$, then
    \begin{equation*}
        A_2 g = \begin{pmatrix} A & I_n \\ 0_{n, n} & 0_{n, n} \end{pmatrix} \begin{pmatrix} 0 \\ a \end{pmatrix} = \begin{pmatrix} a \\ 0 \end{pmatrix}.
    \end{equation*}
    Hence, $A_2 g \in \mathcal{G} \iff a = 0$, i.e., $\mathcal{G}_{A_2} = \{0\}^{2n}$.
    Thus, 
    \begin{equation*}
        \mathcal{V}_2  =  \big\{ v \in  B_2 \mathcal{U} :  v  -  C_2 w \in B_2 \mathcal{U}, \text{for all}\ w \in \mathcal{W} \big\} = \mathcal{Z} \times \{0\}^n,
    \end{equation*}
    because of the architecture of $B_2$ and $C_2$. Then, system \eqref{eq:X Hajek ODE} is related to system \eqref{eq: Hajek} the same way that system \eqref{eq:X ODE split} is related to system \eqref{eq:splitted ODE}.
    Therefore, the following statements are equivalent:
    \begin{itemize}
        \item $x_{tg}$ is resiliently reachable by system \eqref{eq:splitted ODE},
        \item $\mathcal{G}$ is resiliently reachable by system \eqref{eq:X ODE split},
        \item $\mathcal{G}$ is reachable by system \eqref{eq:X Hajek ODE},
        \item $x_{tg}$ is reachable by system \eqref{eq: Hajek}. $\quad \blacksquare$
    \end{itemize}
\end{pf}

Theorem~\ref{thm:Hajek non-zero} transforms resilience of system \eqref{eq:splitted ODE} into bounded controllability of system \eqref{eq: Hajek}, which we verify with Theorem~\ref{thm:Brammer controllability}. 
We can easily adapt the results of Section~\ref{sec: stabilizability} to the resilience case by reusing the same proofs, except that we use Theorems~\ref{thm:Hajek non-zero} and \ref{thm:Brammer controllability} instead of Theorems~\ref{thm:Hajek} and \ref{thm:Brammer stabilizability}.

\begin{prop}\label{prop: resilient reachability Z}
    If $\interior(\mathcal{Z}) \neq \emptyset$, then system~\eqref{eq:splitted ODE} is resilient if and only if $Re(\lambda(A)) = 0$.
\end{prop}

\begin{cor}\label{cor: res control = initial control}
    If $\dim(\mathcal{Z}) = \rank(B)$, then system \eqref{eq:splitted ODE} is resilient if and only if system \eqref{eq:initial ODE} is controllable.
\end{cor}

\begin{thm}[Resilience condition]\label{thm: N&S resilience}
    System \eqref{eq:splitted ODE} is resilient if and only if $\rank\big( \mathcal{C}(A, Z) \big) = n$, \break
    $Re\big(\lambda(A)\big) = 0$, and there is no real eigenvector $v$ of $A^\top$ satisfying $v^\top z \leq 0$ for all $z \in \mathcal{Z}$.
\end{thm}

We now have all the results necessary to solve Problem~\ref{prob:resilience}. However, the condition $Re\big(\lambda(A)\big) = 0$ in Theorem~\ref{thm: N&S resilience} is not satisfied by most systems, that are hence not resilient. This reasoning led us to Problem~\ref{prob:reachable set}, i.e., the determination of the resiliently reachable set of system \eqref{eq:splitted ODE}. Following Theorem~\ref{thm:Hajek non-zero}, we will now study the reachable set of system \eqref{eq: Hajek} given by
\begin{equation*}
    R(T, x_0) := \left\{ e^{AT} \left( x_0 + \int_0^T e^{-At} z(t)\, dt \right), \quad \text{with}\ z(t) \in \mathcal{Z}\ \text{for all}\ t \in [0,T] \right\}.
\end{equation*}
Because analytical study of $R(T,x_0)$ is difficult, most of the research tries to approximate it (see \cite{Girard} and references therein). We want inner approximations of $R(T,x_0)$ in order to determine the states that are guaranteed to be resiliently reachable. We will then present a method of \textit{zonotopic} underapproximation of $R(T,x_0)$ combining the approaches of \cite{Girard} and \cite{Zonotopes}. 

\begin{defn}\label{def: zonotope}
     A \emph{zonotope} $\mathcal{S} \subseteq \mathbb{R}^n$ is a set parametrized by a center $c \in \mathbb{R}^n$ and generators $g_1, \hdots, g_q \in \mathbb{R}^n$ expressed as $\mathcal{S} := \left\{ c + \sum_{i = 1}^q \alpha_i g_i : \alpha_i \in [-1,1] \right\}$ 
    and is denoted $\mathcal{S} = (c, g_1, \hdots g_q)$.
\end{defn}

Note that $B\mathcal{U}$ is a zonotope of center $0$ and generators $B_i$, the columns of $B$. Similarly, $C\mathcal{W} = (0, C_1, \hdots, C_p)$. However, $\mathcal{Z}$ is not a zonotope in general since these sets are not closed under Minkowski difference except for some specific scenarios, as detailed in \cite{Zonotopes}.

Following the work \cite{Zonotopes}, we build an underapproximation of $\mathcal{Z}$ with a symmetric zonotope $\big( 0, g_1, \hdots, g_r \big) \subseteq \mathcal{Z}$ by removing or contracting the generators of $B\mathcal{U}$. We apply the method described in \cite{Girard} to compute efficiently an inner approximation of $R(T, x_0)$. For $N \in \mathbb{N}$, $N \geq 1$, we define 
\begin{equation*}
    \delta t := \frac{T}{N},\quad \Omega_0 := \{ x_0 \}, \quad V := \left\{ \int_0^{\delta t} e^{A(\delta t - t)} z(t)\, dt : z(t) \in \mathcal{Z}\ \text{for}\ t \in [0, \delta t] \right\},
\end{equation*}
 and the recursion $\Omega_{i+1} := e^{A \delta t} \Omega_i \oplus V$. Note that $\Omega_i$ is the exact reachable set $R(i\, \delta t, x_0)$.

However, $V$ is not a zonotope and cannot be computed exactly. Thus, we define the zonotope
\begin{equation*}
    \tilde{V} := \left( 0, \int_0^{\delta t} e^{A(\delta t - t)} g_1\, dt, \hdots, \int_0^{\delta t} e^{A(\delta t - t)} g_r\, dt \right),
\end{equation*}
and $\tilde{V} \subseteq V$ since $\tilde{V}$ corresponds to piecewise constant components of $z(t)$ in $\big( 0, g_1, \hdots, g_r \big)$.

Then, we build $\tilde{\Omega}_0 = \Omega_0 = \{x_0\}$ and $\tilde{\Omega}_{i+1} := e^{A \delta t} \tilde{\Omega}_i \oplus \tilde{V}$, which yields $\tilde{\Omega}_i \subseteq \Omega_i$ for all $i \geq 0$. Since linear maps and Minkowski sums are straightforward on zonotopes \cite{Zonotopes, Girard}, $\tilde{\Omega}_i$ is an easily computable inner approximation of the reachable set $R(i\, \delta t, x_0)$. Note that the precision of the approximation increases with $N$.

Before implementing this solution to Problem~\ref{prob:reachable set} in Section~\ref{subsec: ADMIRE}, we need to answer Problem~\ref{prob:r_q} by quantifying the resilience of linear systems.

\section{Quantitative Resilience}\label{sec:quantitative}

Let us now investigate more complex missions where the target needs to be reached by a certain time. In such scenarios it is crucial to evaluate the maximal time penalty incurred by the malfunctioning system.

Unlike in the driftless case \cite{SIAM_CT}, the optimal reach times $T_N^*$ \eqref{eq: def T_N^*} and $T_M^*$ \eqref{eq: def T_M^*} cannot be reduced to a linear optimization and elude analytical expressions \cite{Athans}. 
Following \cite{Eaton} and \cite{Sakawa} we could numerically compute these reach times, but not the quantitative resilience $r_q$ \eqref{eq:r_q} since it would require computing $T_N^*(x_0)$ and $T_M^*(x_0)$ for all $x_0 \in \mathbb{R}^n$.
Instead, using Lyapunov theory \cite{Kalman}, we establish analytical bound on these two reach times for the target $x_{tg} = 0$ and analytically approximate $r_q$.

\subsection{Nominal reach time}

Assume that $A$ is Hurwitz. Then, for any $Q \succ 0$ there exists $P \succ 0$ such that $PA + A^\top P = -Q$ \cite{Kalman}. Let us consider any such pair $(P,Q)$. We define the Lyapunov function $V(x) := x^\top P x = \|x\|_P^2$ \cite{Khalil}.
Then, for $x$ following \eqref{eq:initial ODE} we have
\begin{equation}\label{eq:Lyapunov}
    \dot V(x) = \dot x^\top P x + x^\top P \dot x = x^\top (A^\top P + PA)x + 2 x^\top P\bar{B}\bar{u} = -x^\top Q x + 2 x^\top P\bar{B}\bar{u}.
\end{equation}
We will now bound $T_N^*(x_0)$.

\begin{prop}\label{prop: T_N lb}
    If system \eqref{eq:initial ODE} is stabilizable and $A$ is Hurwitz, then
    \begin{equation}\label{eq: T_N lb}
        T_N^*(x_0)\ \geq\ 2\frac{\lambda_{min}^P}{\lambda_{max}^Q}\ln \Bigg(1 + \frac{\lambda_{max}^Q \|x_0\|_P}{2 \lambda_{min}^P b_{max}^P }\Bigg),
    \end{equation}
    with $b_{max}^P := \max\big\{ \|\bar{B}\bar{u}\|_P : \bar{u} \in \bar{\mathcal{U}}\big\}$.
\end{prop}
\begin{pf}
    Because $\bar{\mathcal{U}}$ is compact and convex, and system \eqref{eq:initial ODE} is stabilizable, there exists a time-optimal control signal $\bar{u}^* \in \mathcal{F}(\bar{\mathcal{U}})$ driving the state from $x_0$ to the origin in a finite time $T_N^*(x_0)$ \cite{Liberzon}.
    
    We now bound $\dot V$ using \eqref{eq:Lyapunov}. Since $P \succ 0$, there exists $M \in \mathbb{R}^{n \times n}$ such that $P = M^\top M$ \cite{matrix_computations}. Then, $x^\top P \bar{B}\bar{u} = (Mx)^\top M \bar{B} \bar{u} \geq - \|Mx\|_2 \|M\bar{B} \bar{u}\|_2$, by the Cauchy-Schwarz inequality \cite{matrix_computations}. 
    Notice $\|Mx\|_2^2 = x^\top M^\top M x = x^\top P x = \|x\|_P^2$. Similarly, $\|M \bar{B} \bar{u}\|_2 = \|\bar{B} \bar{u}\|_P$.
    
    The maximum $b_{max}^P$ exists since $\bar{\mathcal{U}}$ is compact and the map $\bar{u} \mapsto \|\bar{B}\bar{u}\|_P$ is continuous. Since $Q \succ 0$, we have $x^\top Qx \leq \lambda_{max}^Q \|x\|_2^2$ and $\|x\|_2^2 \leq \|x\|_P^2 / \lambda_{min}^P$ because $P \succ 0$. For $x \neq 0$, we have now lower bounded \eqref{eq:Lyapunov}
    \begin{equation}\label{eq:V dot lb}
        \dot V(x) = \frac{d}{dt} \|x\|_P^2 \geq -\frac{\lambda_{max}^Q}{\lambda_{min}^P} \|x\|_P^2 - 2 b_{max}^P \|x\|_P.
    \end{equation}
    Let $y(t) := \|x(t)\|_P$, $\alpha := \frac{\lambda_{max}^Q}{2\lambda_{min}^P} > 0$, and $\beta := b_{max}^P > 0$. For $x \neq 0$ we divide \eqref{eq:V dot lb} by $2y > 0$ so that $\dot y \geq f(y) := -\alpha y - \beta$. 
    The solution of the differential equation $\dot s(t) = f\big( s(t) \big)$ with $s(0) = y(0)$ is given by $s(t) = e^{-\alpha t}\left( y(0) + \frac{\beta}{\alpha} \right) - \frac{\beta}{\alpha}$. 
    
    Since $f$ is Lipschitz, we can apply the comparison lemma of \cite{Khalil} and we obtain $y(t) \geq s(t)$ for all $t \geq 0$. At time $T = \frac{1}{\alpha} \ln\left(1 + \frac{\alpha}{\beta} y(0) \right)$, we have $s(T) = 0$. Because $\|x(t)\|_P \geq s(t) > 0$ for all $t \in [0, T]$, we have $T_N^*(x_0) \geq T$. Substituting $\alpha$ and $\beta$ yields \eqref{eq: T_N ub}. $\quad \blacksquare$
\end{pf}

The proof of Propositions~\ref{prop: T_N lb}, as well as subsequent Propositions \ref{prop: T_N ub}, \ref{prop: T_M lb}, and \ref{prop: T_M ub}, is shorter than presented in the conference paper \cite{ECC} due to our use of the comparison lemma \cite{Khalil}.
We now upper bound $T_N^*(x_0)$. 

\begin{prop}\label{prop: T_N ub}
    If $\rank(\bar{B}) = n$ and $A$ is Hurwitz, then 
    \begin{equation}\label{eq: T_N ub}
        T_N^*(x_0)\ \leq\ 2\frac{\lambda_{max}^P}{\lambda_{min}^Q}\ln \Bigg(1 + \frac{\lambda_{min}^Q \|x_0\|_P}{2 \lambda_{max}^P b_{min}^P }\Bigg),
    \end{equation}
    with $b_{min}^P := \min\big\{ \| \bar{B}\bar{u} \|_P : \bar{u} \in \partial\bar{\mathcal{U}} \big\}$.
\end{prop}
\begin{pf}
    The minimum $b_{min}^P$ exists since map $\bar{u} \hspace{-0.3mm} \mapsto \hspace{-0.5mm} \|\bar{B}\bar{u}\|_P$ is continuous and $\partial \bar{\mathcal{U}}$ is compact. 
    
    Because $\rank(\bar{B}) = n$, we can choose $\bar{u} \in \mathcal{F}(\bar{\mathcal{U}})$ such that $\bar{B}\bar{u}(t) = -\frac{x(t)}{\|x(t)\|_P} b_{min}^P$ for $x(t) \neq 0$.
    Indeed, assume for contradiction purposes that for some $\tau \geq 0$, $\bar{u}(\tau) \notin \bar{\mathcal{U}}$, i.e., $\|\bar{u}(\tau)\|_\infty > 1$. Let $\hat{u} := \frac{\bar{u}(\tau)}{\|\bar{u}(\tau)\|_\infty}$. Then, $\| \hat{u} \|_\infty = 1$, so $\hat{u} \in \partial \bar{\mathcal{U}}$, but $\| \bar{B} \hat{u}\|_P = \frac{\|\bar{B}\bar{u}(\tau)\|_P}{\|\bar{u}(\tau)\|_\infty} = \frac{b_{min}^P}{\|\bar{u}\|_\infty} < b_{min}^P$, which is a contradiction. Hence, the proposed control signal is admissible and we implement it in \eqref{eq:Lyapunov}.
    
    We obtain $2x^\top P\bar{B}\bar{u} = -2 b_{min}^P \|x\|_P$, so that
    \begin{equation}\label{eq:V dot ub}
        \frac{d}{dt} \|x\|_P^2 = \dot V(x) \leq -\frac{\lambda_{min}^Q}{\lambda_{max}^P} \|x\|_P^2 - 2 b_{min}^P \|x\|_P.
    \end{equation}
    Let $y(t) := \|x(t)\|_P$, $\gamma := \frac{\lambda_{min}^Q}{2\lambda_{max}^P} > 0$, and $\kappa := b_{min}^P > 0$. For $x \neq 0$, dividing \eqref{eq:V dot ub} by $2y > 0$, yields $\dot y \leq f(y) := -\gamma y - \kappa$.
    As in Proposition~\ref{prop: T_N lb}, the comparison lemma of \cite{Khalil} yields $y(t) \leq s(t) = e^{-\gamma t} \left( y(0) + \frac{\kappa}{\gamma} \right) - \frac{\kappa}{\gamma}$ for all $t \geq 0$ as long as $y(t) > 0$. At time $T = \frac{1}{\gamma} \ln\left(1 + \frac{\gamma}{\kappa} y(0) \right)$, $s(T) = 0$. Since $y\big(T_N^*(x_0) \big) = 0$, $T_N^*(x_0) \leq T$. $\quad \blacksquare$
\end{pf}

We now bound the malfunctioning reach time $T_M^*$ following the same method applied to $T_N^*$.

\subsection{Malfunctioning reach time}

We use the same Lyapunov function as above, but with $x$ following \eqref{eq:splitted ODE}, so $\dot V(x) = -x^\top Q x + 2 x^\top P(Bu+Cw)$. We can now lower bound $T_M^*$ as we have done for $T_N^*$.

\begin{prop}\label{prop: T_M lb}
    If system \eqref{eq:splitted ODE} is resiliently stabilizable and $A$ is Hurwitz, then 
    \begin{equation}\label{eq: T_M lb}
        T_M^*(x_0)\ \geq\ 2\frac{\lambda_{min}^P}{\lambda_{max}^Q}\ln \Bigg(1 + \frac{\lambda_{max}^Q \|x_0\|_P}{2 \lambda_{min}^P z_{max}^P }\Bigg),
    \end{equation}
    with $z_{max}^P := \max\big\{ \| z \|_P : z \in \mathcal{Z} \big\}$.
\end{prop}
\begin{pf}
    Since $B\mathcal{U}$ and $C\mathcal{W}$ are compact, $\mathcal{Z}$ is compact \cite{Pontryagin_difference}, so $z_{max}^P$ exists. Since system \eqref{eq:splitted ODE} is resiliently stabilizable, $T_M^*(x_0)$ exists. Let $w^* \in \mathcal{F}(\mathcal{W})$ and $u^* \in \mathcal{F}(\mathcal{W})$ be the arguments of the optimizations in \eqref{eq: def T_M^*}.
    By definition of $\mathcal{Z}$, $z = Cw^* + Bu^* \in \mathcal{F}(\mathcal{Z})$. Then, $\|Cw^*(t) + Bu^*(t)\|_P \leq z_{max}^P$, which yields 
   \begin{equation*}
       \dot V(x) \geq -\frac{\lambda_{max}^Q}{\lambda_{min}^P} \|x\|_P^2 - 2 z_{max}^P \|x\|_P.
   \end{equation*}
    We now proceed as in the second half of the proof of Proposition~\ref{prop: T_N lb} to obtain \eqref{eq: T_M lb}. $\quad \blacksquare$
\end{pf}

Similarly, we upper bound the malfunctioning reach time.

\begin{prop}\label{prop: T_M ub}
    If $\interior(\mathcal{Z}) \neq \emptyset$ and $A$ is Hurwitz, then
    \begin{equation}\label{eq: T_M ub}
        T_M^*(x_0)\ \leq\ 2\frac{\lambda_{max}^P}{\lambda_{min}^Q}\ln \Bigg(1 + \frac{\lambda_{min}^Q \|x_0\|_P}{2 \lambda_{max}^P z_{min}^P }\Bigg),
    \end{equation}
    with $z_{min}^P := \min\big\{ \| z \|_P : z \in \partial \mathcal{Z} \big\}$.
\end{prop}
\begin{pf}
    According to Proposition~\ref{prop: resilient stabilizability Z}, system \eqref{eq:splitted ODE} is resiliently stabilizable, hence a finite $T_M^*$ exists.
    
    Since $\mathcal{Z}$ is compact, so is $\partial \mathcal{Z}$, and thus $z_{min}^P$ exists. Because $\interior(\mathcal{Z}) \neq \emptyset$, according to Lemma~\ref{lemma: int Z non-empty iff 0 in Z}, $0 \in \interior(\mathcal{Z})$. Then, the convexity of $\|\cdot\|_P$ yields $\big\{ z \in \mathbb{R}^n : \|x\|_P \leq z_{min}^P\big\} \subseteq \mathcal{Z}$, so $z(t) := \frac{-x(t)}{\|x(t)\|_P} z_{min}^P \in \mathcal{Z}$.
    
    Let $w^* \in \mathcal{F}(\mathcal{W})$ be the argument of the maximum in \eqref{eq: def T_M^*}. Since $z(t) \in \mathcal{Z}$, there exists $u \in \mathcal{F}(\mathcal{U})$ such that $z(t) = Cw^*(t) + Bu(t)$. Then, applying $w^*$ and $u$ leads to an upper bound of $T_M^*$ since $u$ is not necessarily optimal, while $w^*$ is optimal. Hence
    \begin{equation*}
        \dot V(x) \leq -\frac{\lambda_{min}^Q}{\lambda_{max}^P} \|x\|_P^2 - 2 z_{min}^P \|x\|_P.
    \end{equation*}
   We now proceed as in the second half of the proof of Proposition~\ref{prop: T_N ub} to obtain \eqref{eq: T_M ub}. $\quad \blacksquare$
\end{pf}

We can now bound $T_N^*(x_0)/T_M^*(x_0)$ for all $x_0 \in \mathbb{R}^n$ and hence obtain an approximate of quantitative resilience $r_q$ which cannot be done with prior algorithms \cite{Eaton, Sakawa} that only compute a single instance of $T_N^*(x_0)$ or $T_M^*(x_0)$.

\subsection{Bounding quantitative resilience}

If the system's quantitative resilience $r_q$ is bounded by $\gamma \leq r_q$, then in the worst case, the malfunctioning system will take less than $1/\gamma$ times longer than the nominal system to reach the origin from the same initial state.

\begin{thm}\label{thm: lb rq}
    If $\interior(\mathcal{Z}) \neq \emptyset$ and $A$ is Hurwitz, then
    \begin{equation}\label{eq: lb r_q}
        r_q \geq \max\left( \frac{\lambda_{min}^P \lambda_{min}^Q}{\lambda_{max}^P \lambda_{max}^Q},\ \frac{z_{min}^P}{b_{max}^P} \right),
    \end{equation}
    for any $P \succ 0$ and $Q \succ 0$ such that $A^\top P + PA = -Q$.
\end{thm}
\begin{pf}
    According to Proposition~\ref{prop: resilient stabilizability Z}, system \eqref{eq:splitted ODE} is resiliently stabilizable. Since $\interior(\mathcal{Z}) \neq \emptyset$, we have $\dim(\mathcal{Z}) = n$, and $\mathcal{Z} \subseteq B\mathcal{U} \subseteq \mathbb{R}^n$ yields $\rank(B) = n$. According to Corollary~\ref{cor: res stab = initial stab}, system \eqref{eq:initial ODE} is stabilizable, so we can use \eqref{eq: T_N lb} and \eqref{eq: T_M ub}. We define the positive constants $a := \frac{\lambda_{min}^P \lambda_{min}^Q}{\lambda_{max}^P \lambda_{max}^Q}$, $b := \frac{\lambda_{max}^Q}{2\lambda_{min}^P b_{max}^P}$, and $c := \frac{\lambda_{min}^Q}{2\lambda_{max}^P z_{min}^P}$, so that for $x_0 \in \mathbb{R}^n$, $x_0 \neq 0$, \eqref{eq: T_N lb} and \eqref{eq: T_M ub} yield
    \begin{equation*}
        \frac{T_N^*(x_0)}{T_M^*(x_0)} \geq a \frac{\ln(1 + b \|x_0\|_P)}{\ln(1 + c \|x_0\|_P)} := f( \|x_0\|_P).
    \end{equation*}
    Then, according to \eqref{eq:r_q}, $r_q \geq \underset{x_0\, \in\, \mathbb{R}^n}{\inf} f(\|x_0\|_P)$.
    
    If $b = c$, then $f(s) = a$ for all $s \geq 0$, so $r_q \geq a$. If $b > c$, then $f$ is increasing, so $\inf \big\{ f(s) : s > 0\big\} = \underset{s \rightarrow 0}{\lim}\, f(s)$. L'H\^opital's Rule \cite{real_variables} yields
    \begin{equation*}
        \underset{s \rightarrow 0}{\lim}\, f(s) = \underset{s \rightarrow 0}{\lim}\, a \frac{\ln(1 + b s)}{\ln(1 + c s)} = \underset{s \rightarrow 0}{\lim}\, a \frac{ \frac{b}{1 + b s} }{ \frac{c}{1 + c s} } = \frac{ab}{c}.
    \end{equation*}
    Then, $f(0) = \frac{ab}{c} = \frac{z_{min}^P}{b_{max}^P} > a$. 
    If $c > b$, then $f$ is decreasing, so $\inf \big\{ f(s) : s \geq 0\big\} = \underset{s \rightarrow +\infty}{\lim} f(s) = a$ by L'H\^opital's Rule \cite{real_variables}. To sum up, $\underset{s\, \geq\, 0}{\inf} f(s) = \max \big( a,\, \frac{ab}{c} \big) \leq r_q$. $\quad \blacksquare$
\end{pf}

We can upper bound $r_q$ using a similar approach.

\begin{thm}\label{thm: ub rq}
    If $\rank(\bar{B}) = n$, $A$ is Hurwitz, and system \eqref{eq:splitted ODE} is resiliently stabilizable, then
     \begin{equation}\label{eq: ub r_q}
        r_q \leq \max\left( \frac{\lambda_{max}^P \lambda_{max}^Q}{\lambda_{min}^P \lambda_{min}^Q},\ \frac{z_{max}^P}{b_{min}^P} \right),
    \end{equation}
    for any $P \succ 0$ and $Q \succ 0$ such that $A^\top P + PA = -Q$.
\end{thm}
\begin{pf}
    With our assumptions we are allowed to use Propositions~\ref{prop: T_N ub} and \ref{prop: T_M lb}. We define the positive constants $a := \frac{\lambda_{max}^P \lambda_{max}^Q}{\lambda_{min}^P \lambda_{min}^Q}$, $b := \frac{\lambda_{min}^Q}{2\lambda_{max}^P b_{min}^P}$, and $c := \frac{\lambda_{max}^Q}{2\lambda_{min}^P z_{max}^P}$, so that for $x_0 \in \mathbb{R}^n$, $x_0 \neq 0$, \eqref{eq: T_N ub} and \eqref{eq: T_M lb} yield
    \begin{equation*}
        \frac{T_N^*(x_0)}{T_M^*(x_0)} \leq a \frac{\ln(1 + b \|x_0\|_P)}{\ln(1 + c \|x_0\|_P)} := g( \|x_0\|_P).
    \end{equation*}
    Then, according to \eqref{eq:r_q}, $r_q \leq \underset{x_0\, \in\, \mathbb{R}^n}{\inf} g(\|x_0\|_P)$. This function $g$ is similar to $f$ in the proof of Theorem~\ref{thm: lb rq}, and thus $r_q \leq \underset{x_0\, \in\, \mathbb{R}^n}{\inf} g(\|x_0\|_P) = \max \big( a,\, a\frac{b}{c} \big)$, yielding \eqref{eq: ub r_q}. $\quad \blacksquare$
\end{pf}



Theorems~\ref{thm: lb rq} and \ref{thm: ub rq} bound $r_q$ and hence solve Problem~\ref{prob:r_q}. We will now apply the developed theory to two examples.

\section{Numerical Results}\label{sec:results}

We will first study the resilient reachability of the ADMIRE fighter jet model \cite{FOI_Admire}, before quantifying the resilience of a temperature control system.

\subsection{Resilient reachability of the ADMIRE fighter jet model}\label{subsec: ADMIRE}

\begin{figure}[htbp!]
    \centering
    \includegraphics[scale=0.6]{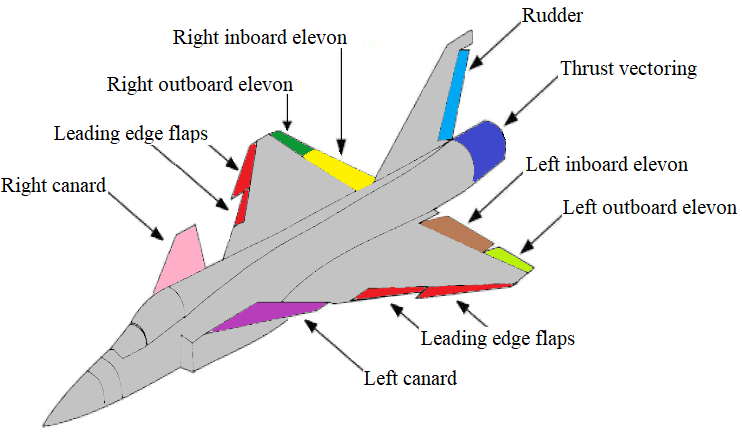}
    \caption{The ADMIRE fighter jet model. Image modified from \cite{FOI_Admire} with a different color for each independent actuator.}
    \label{fig:admire}
\end{figure}

The ADMIRE model has already served as an application case in several control frameworks \cite{ADMIRE_2, TAC} and is illustrated on Fig.~\ref{fig:admire}.

Relying on the simulation package \textit{Admirer4p1}\footnote{\scriptsize \url{https://app.box.com/s/r9wfyjd9o4pq2if9xhd17yxeqc36j7ei}} we run the ADMIRE simulation in MATLAB and obtain the linearized dynamics at Mach $0.3$ and altitude $2000\,m$. We scale $\bar{B}$ so that the input set of each actuator from \cite{FOI_Admire} is scaled to $[-1, 1]$. The states and matrices of the system $\dot X(t) = AX(t) + \bar{B}\bar{u}(t)$ are given below.

Consider a scenario in which, after sustaining damage, an actuator of the fighter jet starts producing uncontrolled and possibly undesirable inputs. By studying $\bar{B}$, we gain intuition on the resilience of the jet. The effect of the yaw (resp. pitch) thrust vectoring on the yaw (resp. pitch) rate is larger than that of all the other actuators combined, which gives the intuition that the jet is not resilient to the loss control over thrust vectoring. None of the other actuators produce such a dominant effect, hence giving the intuition that the jet is resilient to the loss of control over any one of the first eight actuators.

\begin{figure}[htbp!]
    \centering
    \begin{subfigure}[]{0.49\columnwidth}
        \includegraphics[scale = 0.6]{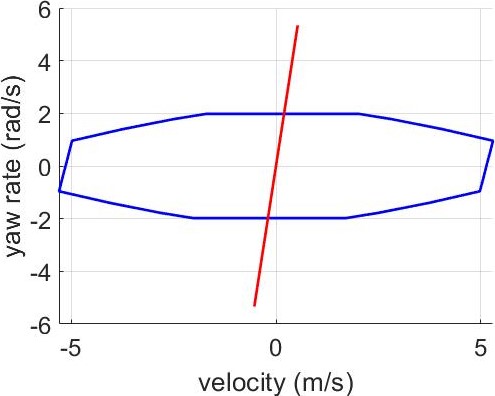}
        \caption{Yaw thrust vectoring.}
        \label{subfig:yaw thrust vectoring}
    \end{subfigure}\hfill
    \begin{subfigure}[]{0.49\columnwidth}
        \includegraphics[scale = 0.6]{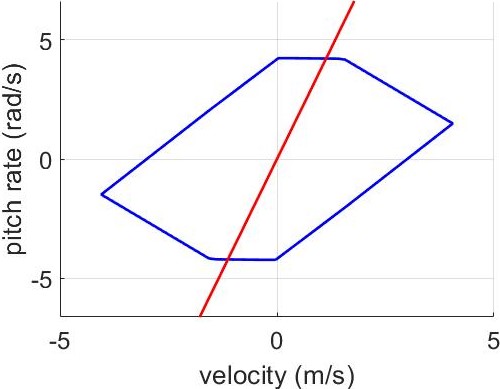}
        \caption{Pitch thrust vectoring.}
        \label{subfig:pitch thrust vectoring}
    \end{subfigure}
    \caption{2D projection of sets $B\mathcal{U}$ (blue) and $C\mathcal{W}$ (red) for the loss of control over the two thrust vectoring actuators.}
    \label{fig:thrust vectoring out}
\end{figure}

Following Lemma~\ref{lemma: not stabilizable}, we test our intuition by verifying whether $C\mathcal{W} \subseteq B\mathcal{U}$. These sets are zonotopes of dimension 9, represented in MATLAB using function $zonotope(\cdot)$ from the CORA package \cite{CORA}. The associated function $in(\cdot)$ is employed to verify their inclusion.
As expected, $C\mathcal{W} \subseteq B\mathcal{U}$ for the loss of control over any one actuator except for the thrust vectoring ones, as shown on Fig.~\ref{fig:thrust vectoring out}. 
Note that for any projection $\proj(\cdot)$, we have $\proj(C\mathcal{W}) \nsubseteq \proj(B\mathcal{U})$ implies $C\mathcal{W} \nsubseteq B\mathcal{U}$, but $\proj(C\mathcal{W}) \subseteq \proj(B\mathcal{U})$ does not yield $C\mathcal{W} \subseteq B\mathcal{U}$.

\begin{figure*}[htbp!]
\centering
\begin{align*}
    X \hspace{-1mm} = \hspace{-1mm} \def\arraystretch{1.1}\begin{pmatrix} v \\ \alpha \\ \beta \\ p \\ q \\ r \\ \psi \\ \theta \\ \varphi \end{pmatrix} \hspace{-1mm} \begin{array}{l} \text{velocity}\ (m/s), \\ \text{angle of attack}\ (rad), \\ \text{sideslip angle}\ (rad), \\ \text{roll rate}\ (rad/s), \\ \text{pitch rate}\ (rad/s), \\ \text{yaw rate}\ (rad/s), \\ \text{heading angle}\ (rad), \\ \text{pitch angle}\ (rad), \\ \text{roll angle}\ (rad), \end{array} \hspace{-1mm} A  \hspace{-0.5mm} = \hspace{-1mm} \left(\arraycolsep=1mm\def\arraystretch{1.1}\begin{array}{ccccccccc}
    -0.02 & -4.65 & 0.37 & 0 & -0.3 & 0 & 0 & -9.81 & 0 \\
    0 & -0.78 & 0.01 & 0 & 0.97 & 0 & 0 & 0 & 0 \\
    0 & 0 & -0.19 & 0.12 & 0 & -0.98 & 0 & 0 & 0.1 \\
    0 & 0 & -15.47 & -1.5 & 0 & 0.54 & 0 & 0 & 0 \\
    0 & 4.18 & -0.01 & 0 & -0.78 & 0 & 0 & 0 & 0 \\
    0 & 0 & 0.95 & -0.09 & 0 & -0.34 & 0 & 0 & 0 \\
    0 & 0 & 0 & 0 & 0 & 1.01 & 0 & 0 & 0 \\
    0 & 0 & 0 & 0 & 1 & 0 & 0 & 0 & 0 \\
    0 & 0 & 0 & 1 & 0 & 0.12 & 0 & 0 & 0 \end{array}\right) \\
    \bar{B}^\top = \left(\arraycolsep=1mm\def\arraystretch{1.1}\begin{array}{ccccccccc}
       -0.62 & 0 & 0 & 0.37 & 0.67 & -0.19 & 0 & 0 & 0 \\
       -0.62 & 0 & 0 & -0.37 & 0.67 & 0.19 & 0 & 0 & 0 \\
       -0.4 & -0.02 & 0 & -2.27 & -0.55 & -0.1 & 0 & 0 & 0 \\
       -0.62 & -0.04 & 0.01 & -1.96 & -0.88 & -0.22 & 0 & 0 & 0 \\
       -0.62 & -0.04 & -0.01 & 1.96 & -0.88 & 0.22 & 0 & 0 & 0 \\
       -0.4 & -0.02 & 0 & 2.27 & -0.55 & 0.1 & 0 & 0 & 0 \\
       -0.16 & 0 & 0.02 & 1.59 & 0 & -0.96 & 0 & 0 & 0 \\
        0.08 & 0 & 0 & 0 & -0.02 & 0 & 0 & 0 & 0 \\
       -0.53 & 0 & 0.11 & -0.64 & 0.01 & -5.34 & 0 & 0 & 0 \\
       -1.78 & -0.11 & 0 & 0 & -6.63 & 0 & 0 & 0 & 0 \end{array}\right)        \def\arraystretch{1.1}\begin{array}{l} \text{right canard,} \\
       \text{left canard,} \\
       \text{right outboard elevon,} \\
       \text{right inboard elevon,} \\
       \text{left inboard elevon,} \\
       \text{left outboard elevon,} \\
       \text{rudder,} \\
       \text{leading edge flaps,} \\
       \text{yaw thrust vectoring,} \\
       \text{pitch thrust vectoring.} \end{array}
\end{align*}
\end{figure*}

The eigenvalues of $A$ do not verify either $Re(\lambda(A)) = 0$ or $Re(\lambda(A)) \leq 0$. Thus, the system is neither resilient nor resiliently stabilizable. However, as anticipated with Problem~\ref{prob:reachable set}, the linearized model is only valid locally and hence we should only study the resilient reachability of targets close to the linearization equilibrium.

We follow the method detailed in Section~\ref{sec: reachability} to approximate the resiliently reachable set of the malfunctioning system.
Assume the pilot lost control over the right outboard elevon $\bar{u}_3$. We use the CORA \cite{CORA} function $minus(\cdot, \cdot)$ to underapproximate the Minkowski difference $\mathcal{Z} = B\mathcal{U} \ominus C\mathcal{W}$ as a zonotope $(0, g_1, \hdots, g_9)$, following the method of \cite{Zonotopes}. We take $T = 0.2\,s$ and $N = 5$.
Then, we underapproximate $R(T, x_0)$ with $\tilde{\Omega}_N$ using the recursion $\tilde{\Omega}_{i+1} = e^{A\delta t} \tilde{\Omega}_i \oplus \tilde{V}$ of Section~\ref{sec: reachability}.

\begin{figure}[htbp!]
    \centering
    \includegraphics[scale = 0.5]{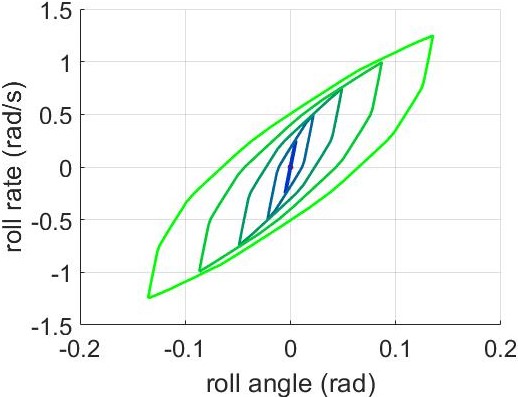}
    \caption{Projection of $\tilde{\Omega}_1, \hdots, \tilde{\Omega}_5$ on the $(\phi, p)$ plane.}
    \label{fig:reachable p phi}
\end{figure}

Since the malfunctioning actuator $\bar{u}_3$ has a strong impact on the roll rate $p$ of the jet, we want to see what range of roll rates is reachable. We compute $\tilde{\Omega}_1, \hdots, \tilde{\Omega}_N$ and project them in 2D as shown on Fig.~\ref{fig:reachable p phi}. Then, in time $T$ the jet can change its roll rate up to $\pm 1.2\,rad/s$, despite the loss of control over the right outboard elevon.

\begin{figure}[htbp!]
    \centering
    \begin{subfigure}[]{0.49\columnwidth}
        \includegraphics[scale = 0.5]{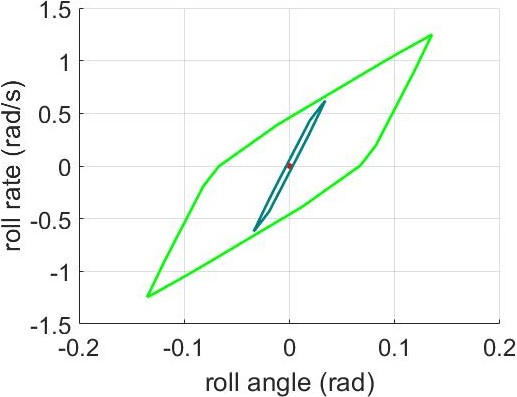}
        \caption{$N = 2$.}
        \label{subfig: N=2}
    \end{subfigure}\hfill
    \begin{subfigure}[]{0.49\columnwidth}
        \includegraphics[scale = 0.5]{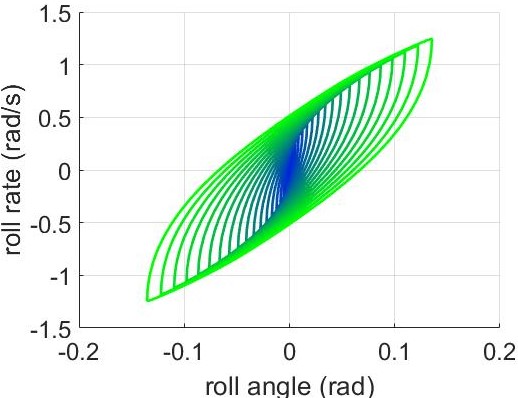}
        \caption{$N = 20$.}
        \label{subfig: N=20}
    \end{subfigure}
    \caption{Projection of $\tilde{\Omega}_1, \hdots, \tilde{\Omega}_N$ on the $(\phi, p)$ plane for different values of $N$.}
    \label{fig: N = 2 and 20}
\end{figure}

We now study the impact of $N$, i.e., of $\delta t$ on the precision of $\tilde{\Omega}_N$ to approximate the real reachable set $R(T,x_0)$ when keeping $T$ constant. Since $\dim \big(R(T,x_0)\big) = 9$, we will only study the impact on the range of roll rates reachable at roll angle $\phi = 0\,rad$.
For $N = 2$ the reachable range of roll rates is $\pm 0.37\,rad/s$, while for $N = 5$ it is $\pm 0.42\,rad/s$, and $\pm 0.43\,rad/s$ for $N = 20$, as illustrated on Fig.~\ref{fig:reachable p phi} and \ref{fig: N = 2 and 20}.
Hence, as explained in Section~\ref{sec: reachability}, increasing $N$ raises nonlinearly the precision of $\tilde{\Omega}_N$ and increases linearly the computational cost since $\tilde{\Omega}_N$ is a zonotope with $9N$ generators.

Now assume that the in-flight damage responsible for the loss of control over the elevon $\bar{u}_3$ also initially caused it to jerk resulting in a sudden jump in roll rate. Then, instead of $X(0) = 0$ we have $p(0) = 0.44\,rad/s$ and the goal is to stabilize the jet at the origin $X_{tg}$. 

\begin{figure}[htbp!]
    \centering
    \includegraphics[scale = 0.5]{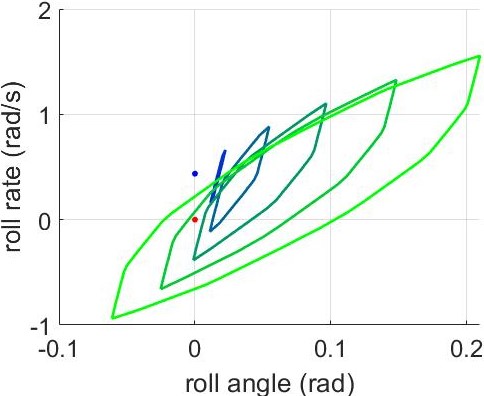}
    \caption{Projection of $\tilde{\Omega}_1, \hdots, \tilde{\Omega}_5$ on the $(\phi, p)$ plane. Initial state $X_0$ is the blue dot, target $X_{tg}$ is the red dot, and $N = 5$.}
    \label{fig:p phi 25deg/s}
\end{figure}

We can see on Fig.~\ref{fig:p phi 25deg/s} that the target only enters the projection of the reachable set after 4 iterations of $\delta t = 0.04\,s$, i.e., for $t \geq 0.16\,s$. By choosing a smaller $\delta t$ we can refine the precision on the minimal entering time. However, to calculate the reachable time $T_M^*(X_0, X_{tg})$ we need to use the CORA function $in(\cdot)$ to verify whether $X_{tg} \in \tilde{\Omega}_N$ since Fig.~\ref{fig:p phi 25deg/s} is only a 2D projection of the 9D reachable set and could be deceiving. Indeed, for $p(0) = 0.5\,rad/s$, the 2D projection is similar to Fig.~\ref{fig:p phi 25deg/s} with the red dot inside the projection of $\tilde{\Omega}_N$, but $X_{tg} \notin \tilde{\Omega}_N$.

We successfully demonstrated the developed resilience theory and the zonotopic method to underapproximate the resiliently reachable set of the ADMIRE jet model.

\subsection{Temperature control system}\label{subsec: temperature}

We now illustrate our quantitative resilience bounds on a temperature control system motivated by \cite{temperature} and illustrated on Fig.~\ref{fig: n,m rooms}.

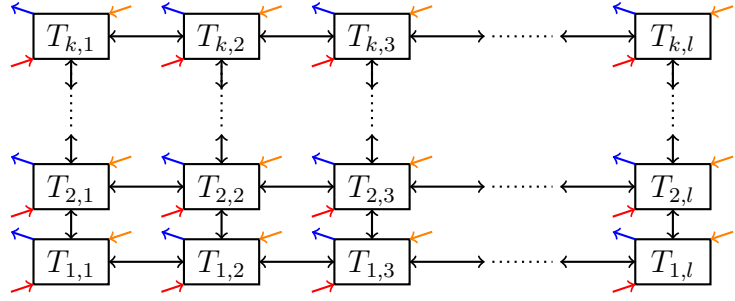
\begin{figure}[htbp!]
    \centering
    \begin{tikzpicture}[scale = 1]
        \draw[thick] (0, 0) -- (1, 0) -- (1, 0.6) -- (0, 0.6) -- (0, 0);
        \node at (0.5, 0.25) {$T_{1,1}$};
        \draw[red, thick, ->] (-0.3, -0.1) -- (0, 0);
        \draw[blue, thick, <-] (-0.3, 0.7) -- (0, 0.6);
        \draw[orange, thick, ->] (1.3, 0.7) -- (1, 0.6);
        \draw[thick, <->] (1, 0.3) -- (2, 0.3);
        \draw[thick, <->] (0.5, 0.6) -- (0.5, 1);
        
        \draw[thick] (2, 0) -- (3, 0) -- (3, 0.6) -- (2, 0.6) -- (2, 0);
        \node at (2.5, 0.25) {$T_{1,2}$};
        \draw[red, thick, ->] (1.7, -0.1) -- (2, 0);
        \draw[blue, thick, <-] (1.7, 0.7) -- (2, 0.6);
        \draw[orange, thick, ->] (3.3, 0.7) -- (3, 0.6);
        \draw[thick, <->] (3, 0.3) -- (4, 0.3);
        \draw[thick, <->] (2.5, 0.6) -- (2.5, 1);
        
        \draw[thick] (4, 0) -- (5, 0) -- (5, 0.6) -- (4, 0.6) -- (4, 0);
        \node at (4.5, 0.25) {$T_{1,3}$};
        \draw[red, thick, ->] (3.7, -0.1) -- (4, 0);
        \draw[blue, thick, <-] (3.7, 0.7) -- (4, 0.6);
        \draw[orange, thick, ->] (5.3, 0.7) -- (5, 0.6);
        \draw[thick, <->] (5, 0.3) -- (6, 0.3);
        \draw[thick, <->] (4.5, 0.6) -- (4.5, 1);
        
        \draw[dotted, thick] (6.1, 0.3) -- (6.9, 0.3);
        
        \draw[thick] (8, 0) -- (9, 0) -- (9, 0.6) -- (8, 0.6) -- (8, 0);
        \node at (8.5, 0.25) {$T_{1,l}$};
        \draw[red, thick, ->] (7.7, -0.1) -- (8, 0);
        \draw[blue, thick, <-] (7.7, 0.7) -- (8, 0.6);
        \draw[orange, thick, ->] (9.3, 0.7) -- (9, 0.6);
        \draw[thick, <->] (7, 0.3) -- (8, 0.3);
        \draw[thick, <->] (8.5, 0.6) -- (8.5, 1);
        
        \draw[thick] (0, 1) -- (1, 1) -- (1, 1.6) -- (0, 1.6) -- (0, 1);
        \node at (0.5, 1.25) {$T_{2,1}$};
        \draw[red, thick, ->] (-0.3, 0.9) -- (0, 1);
        \draw[blue, thick, <-] (-0.3, 1.7) -- (0, 1.6);
        \draw[orange, thick, ->] (1.3, 1.7) -- (1, 1.6);
        \draw[thick, <->] (1, 1.3) -- (2, 1.3);
        \draw[thick, <->] (0.5, 1.6) -- (0.5, 2);
        
        \draw[thick] (2, 1) -- (3, 1) -- (3, 1.6) -- (2, 1.6) -- (2, 1);
        \node at (2.5, 1.25) {$T_{2,2}$};
        \draw[red, thick, ->] (1.7, 0.9) -- (2, 1);
        \draw[blue, thick, <-] (1.7, 1.7) -- (2, 1.6);
        \draw[orange, thick, ->] (3.3, 1.7) -- (3, 1.6);
        \draw[thick, <->] (3, 1.3) -- (4, 1.3);
        \draw[thick, <->] (2.5, 1.6) -- (2.5, 2);
        
        \draw[thick] (4, 1) -- (5, 1) -- (5, 1.6) -- (4, 1.6) -- (4, 1);
        \node at (4.5, 1.25) {$T_{2,3}$};
        \draw[red, thick, ->] (3.7, 0.9) -- (4, 1);
        \draw[blue, thick, <-] (3.7, 1.7) -- (4, 1.6);
        \draw[orange, thick, ->] (5.3, 1.7) -- (5, 1.6);
        \draw[thick, <->] (5, 1.3) -- (6, 1.3);
        \draw[thick, <->] (4.5, 1.6) -- (4.5, 2);
        
        \draw[dotted, thick] (6.1, 1.3) -- (6.9, 1.3);
        
        \draw[thick] (8, 1) -- (9, 1) -- (9, 1.6) -- (8, 1.6) -- (8, 1);
        \node at (8.5, 1.25) {$T_{2,l}$};
        \draw[red, thick, ->] (7.7, 0.9) -- (8, 1);
        \draw[blue, thick, <-] (7.7, 1.7) -- (8, 1.6);
        \draw[orange, thick, ->] (9.3, 1.7) -- (9, 1.6);
        \draw[thick, <->] (7, 1.3) -- (8, 1.3);
        \draw[thick, <->] (8.5, 1.6) -- (8.5, 2);
        
        \draw[dotted, thick] (0.5, 2.1) -- (0.5, 2.9);
        \draw[dotted, thick] (2.5, 2.1) -- (2.5, 2.9);
        \draw[dotted, thick] (4.5, 2.1) -- (4.5, 2.9);
        \draw[dotted, thick] (8.5, 2.1) -- (8.5, 2.9);
        
        \draw[thick] (0, 3) -- (1, 3) -- (1, 3.6) -- (0, 3.6) -- (0, 3);
        \node at (0.5, 3.25) {$T_{k,1}$};
        \draw[red, thick, ->] (-0.3, 2.9) -- (0, 3);
        \draw[blue, thick, <-] (-0.3, 3.7) -- (0, 3.6);
        \draw[orange, thick, ->] (1.3, 3.7) -- (1, 3.6);
        \draw[thick, <->] (1, 3.3) -- (2, 3.3);
        \draw[thick, <->] (0.5, 2.6) -- (0.5, 3);
        
        \draw[thick] (2, 3) -- (3, 3) -- (3, 3.6) -- (2, 3.6) -- (2, 3);
        \node at (2.5, 3.25) {$T_{k,2}$};
        \draw[red, thick, ->] (1.7, 2.9) -- (2, 3);
        \draw[blue, thick, <-] (1.7, 3.7) -- (2, 3.6);
        \draw[orange, thick, ->] (3.3, 3.7) -- (3, 3.6);
        \draw[thick, <->] (3, 3.3) -- (4, 3.3);
        \draw[thick, <->] (2.5, 2.6) -- (2.5, 3);
        
        \draw[thick] (4, 3) -- (5, 3) -- (5, 3.6) -- (4, 3.6) -- (4, 3);
        \node at (4.5, 3.25) {$T_{k,3}$};
        \draw[red, thick, ->] (3.7, 2.9) -- (4, 3);
        \draw[blue, thick, <-] (3.7, 3.7) -- (4, 3.6);
        \draw[orange, thick, ->] (5.3, 3.7) -- (5, 3.6);
        \draw[thick, <->] (5, 3.3) -- (6, 3.3);
        \draw[thick, <->] (4.5, 2.6) -- (4.5, 3);
        
        \draw[dotted, thick] (6.1, 3.3) -- (6.9, 3.3);
        
        \draw[thick] (8, 3) -- (9, 3) -- (9, 3.6) -- (8, 3.6) -- (8, 3);
        \node at (8.5, 3.25) {$T_{k,l}$};
        \draw[red, thick, ->] (7.7, 2.9) -- (8, 3);
        \draw[blue, thick, <-] (7.7, 3.7) -- (8, 3.6);
        \draw[orange, thick, ->] (9.3, 3.7) -- (9, 3.6);
        \draw[thick, <->] (7, 3.3) -- (8, 3.3);
        \draw[thick, <->] (8.5, 2.6) -- (8.5, 3);
        
    \end{tikzpicture}
    \caption{Heat exchange graph of an office building with $k$ floors of $l$ rooms, each at a temperature $T_{i,j}$.}
    \label{fig: n,m rooms}
\end{figure}

We study a scenario where a worker remains in their office after hours and manually opens or closes their door and window, thus overriding the building heat controller which aims at maintaining a target temperature $T_{tg}$. After this loss of control, we will compare our analytical bounds on the nominal and malfunctioning reach times with the numerical results of \cite{Eaton, Sakawa}. We will also bound the quantitative resilience of the system which could not be done with prior work and motivated the analytical bounds of Section~\ref{sec:quantitative}.

The controller uses a central heater $q_h$, central AC $q_{AC}$, and incrementally opens doors $q_d$ and windows $q_w$ for room specific adjustments. The controller also takes advantage of solar heating $q_S$, heat losses through the outside wall $q_l$, and heat transfers between adjoining rooms $q_{adj}$. The temperature dynamics are then
\begin{equation*}
    mC_p \dot T_{i,j} = q_h \hspace{-0.5mm} - q_{AC} + q_{d_{i,j}} \hspace{-1mm} - q_{w_{i,j}} + q_{S_{i,j}} \hspace{-1mm} - q_{l_{i,j}} + \sum q_{adj}
\end{equation*}
with $m$ the mass of air in each room, $C_p$ its specific heat capacity, $q_{adj} = aU(T_{adj} - T_{i,j})$, with $a$ the area of the wall between rooms, and $U$ the overall heat transfer coefficient between adjoining rooms, which depends on the wall materials. 
To have symmetric inputs, we combine the heat transfers in pairs: $q_h - q_{AC} =: Q_{hAC} u_{hAC}$, $q_{d_{i,j}} - q_{w_{i,j}} =: Q_{dw} u_{dw}^{i,j}$, and $q_{S_{i,j}} - q_{l_{i,j}} =: Q_{Sl} u_{Sl}^{i,j}$ with $u_{hAC}$, $u_{dw}^{i,j}$, and $u_{Sl}^{i,j} \in [-1, 1]$.

\begin{figure}[htbp!]
    \centering
    \begin{tikzpicture}[scale = 1]
        \draw[dotted, thin] (-3, 0) -- (3,0);
        \draw[dotted, thin] (-3, 2) -- (3,2);
        \draw[very thick] (-4.5, 2) -- (-3, 2);
        \draw[very thick] (3, 2) -- (4.5,2);
        \draw[very thick] (-4.5, 0) -- (-3, 0);
        \draw[very thick] (3, 0) -- (4.5,0);
        \node at (-4, 1) {$T_{tg}$};
        \node at (4, 1) {$T_{tg}$};
        \draw[very thick] (-4.5, -0.6) -- (4.5, -0.6);
        \node at (3.8, -0.3) {\textcolor{red}{hallway}};
        
        \node at (4, 2.8) {\textcolor{blue}{outside}};
        \node at (-4.2, 2.8) {\textcolor{orange}{Sun}};
        
        \draw[very thick] (-2.3, 0) -- (-3, 0) -- (-3, 2) -- (-2.3, 2) -- (-2.1, 2.2);
        \draw[very thick] (-1.9, 2.2) -- (-1.7, 2) -- (-1, 2) -- (-1, 0) -- (-1.7, 0) -- (-2.1, -0.4);
        \node at (-2, 1) {$T_1$};
        \draw[blue, ultra thick, <-] (-2, 2.6) -- (-2, 1.8);
        \node at (-2, 2.8) {\textcolor{blue}{$q_{w1}$}};
        \draw[red, ultra thick, ->] (-2.3, -0.2) -- (-1.8, 0.3);
        \node at (-2.6, -0.3) {\textcolor{red}{$q_{d1}$}};
        \draw[ultra thick, <->] (-3.3, 1) -- (-2.7, 1);
        \node at (-2.6, 0.6) {\textcolor{black}{$q_{g1}$}};
        \draw[orange, ->] (-3.2, 2.5) -- (-2.7, 2);
        \draw[orange, ->] (-3.3, 2.5) -- (-2.8, 2);
        \draw[orange, ->] (-3.1, 2.5) -- (-2.6, 2);
        \node at (-3.2, 2.8) {\textcolor{orange}{$q_{S1}$}};
        \draw[blue, ->] (-1.4, 1.8) -- (-1.4, 2.2);
        \draw[blue, ->] (-1.6, 1.8) -- (-1.6, 2.2);
        \draw[blue, ->] (-1.2, 1.8) -- (-1.2, 2.2);
        \node at (-1.4, 1.6) {\textcolor{blue}{$q_{l1}$}};

        \draw[very thick] (-0.3, 0) -- (-1, 0) -- (-1, 2) -- (-0.3, 2) -- (-0.1, 2.2);
        \draw[very thick] (0.1, 2.2) -- (0.3, 2) -- (1, 2) -- (1, 0) -- (0.3, 0) -- (-0.1, -0.4);
        \node at (0, 1) {$T_2$};
        \draw[blue, ultra thick, <-] (0, 2.6) -- (0, 1.8);
        \node at (0, 2.8) {\textcolor{blue}{$q_{w2}$}};
        \draw[red, ultra thick, ->] (-0.3, -0.2) -- (0.2, 0.3);
        \node at (-0.6, -0.3) {\textcolor{red}{$q_{d2}$}};
        \draw[ultra thick, <->] (-1.3, 1) -- (-0.7, 1);
        \node at (-0.6, 0.6) {\textcolor{black}{$q_{12}$}};
        \draw[orange, ->] (-1, 2.5) -- (-0.5, 2);
        \draw[orange, ->] (-1.1, 2.5) -- (-0.6, 2);
        \draw[orange, ->] (-0.9, 2.5) -- (-0.4, 2);
        \node at (-1, 2.8) {\textcolor{orange}{$q_{S2}$}};
        \draw[blue, ->] (0.4, 1.8) -- (0.4, 2.2);
        \draw[blue, ->] (0.6, 1.8) -- (0.6, 2.2);
        \draw[blue, ->] (0.8, 1.8) -- (0.8, 2.2);
        \node at (0.6, 1.6) {\textcolor{blue}{$q_{l2}$}};

        \draw[very thick] (1.7, 0) -- (1, 0) -- (1, 2) -- (1.7, 2) -- (1.9, 2.2);
        \draw[very thick] (2.1, 2.2) -- (2.3, 2) -- (3, 2) -- (3, 0) -- (2.3, 0) -- (1.9, -0.4);
        \node at (2, 1) {$T_3$};
        \draw[blue, ultra thick, <-] (2, 2.6) -- (2, 1.8);
        \node at (2, 2.8) {\textcolor{blue}{$q_{w3}$}};
        \draw[red, ultra thick, ->] (1.7, -0.2) -- (2.2, 0.3);
        \node at (1.4, -0.3) {\textcolor{red}{$q_{d3}$}};
        \draw[ultra thick, <->] (3.3, 1) -- (2.7, 1);
        \node at (1.4, 0.6) {\textcolor{black}{$q_{23}$}};
        \draw[ultra thick, <->] (1.3, 1) -- (0.7, 1);
        \node at (3.4, 0.6) {\textcolor{black}{$q_{3g}$}};
        \draw[orange, ->] (1, 2.5) -- (1.5, 2);
        \draw[orange, ->] (1.1, 2.5) -- (1.6, 2);
        \draw[orange, ->] (0.9, 2.5) -- (1.4, 2);
        \node at (1, 2.8) {\textcolor{orange}{$q_{S3}$}};
        \draw[blue, ->] (2.4, 1.8) -- (2.4, 2.2);
        \draw[blue, ->] (2.6, 1.8) -- (2.6, 2.2);
        \draw[blue, ->] (2.8, 1.8) -- (2.8, 2.2);
        \node at (2.6, 1.6) {\textcolor{blue}{$q_{l3}$}};
        
    \end{tikzpicture}
    \caption{Scheme of the rooms and of the heat transfers. The heater $q_h$ and AC transfers $q_{AC}$ are not shown for clarity.}
    \label{fig:rooms}
\end{figure}
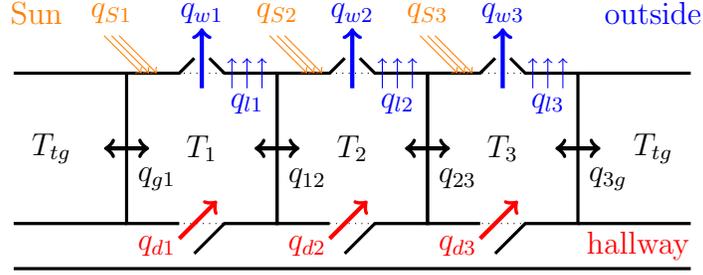

We write the dynamics as $\dot T = AT + \bar{B}\bar{u}$, with
\begin{equation*}
    A = \frac{a}{m C_p} \begin{pmatrix} -2U & U & 0 & 0 & \hdots & 0 & U & 0 & 0 & \hdots \\
                                        U & -3U & U & 0 & \hdots & 0 & 0 & U & 0 & \hdots\\
                                        0 & \ddots & & \ddots & & \ddots & & \ddots & & \ddots \end{pmatrix},
\end{equation*}
\begin{equation*}
    \bar{B} = \frac{1}{m C_p}\begin{pmatrix} Q_{Sl} I_{kl, kl} & Q_{dw} I_{kl, kl} & Q_{hAC} \mathbf{1}_{kl} \end{pmatrix},
\end{equation*}
$\bar{u}^\top = \big( u_{Sl}^{1,1}, \hdots, u_{Sl}^{k,l}, u_{dw}^{1,1}, \hdots, u_{dw}^{k,l}, u_{hAC} \big) \in \mathbb{R}^{2kl+1}$ and $T^\top = \big(T_{1,1}, \hdots, T_{k,l}\big) \in \mathbb{R}^{kl}$. To perform numerical calculations, we restrict our building to $k = 1$ and $l = 3$, as schematized in Fig.~\ref{fig:rooms}.

Taking $x := T - T_{tg}$, the heat dynamics of the system illustrated on Fig.~\ref{fig:rooms} are $\dot x = Ax + \bar{B}\bar{u}$ with $x_{tg} = 0$ and
\begin{equation*}
    A = \frac{a}{m C_p} \begin{pmatrix} -U_{g1} -U_{12} & U_{12} & 0 \\ U_{12} & -U_{12} - U_{23} & U_{23} \\ 0 & U_{23} & - U_{23} - U_{3g} \end{pmatrix}.
\end{equation*}

Based on \cite{temperature}, we use the following values: $a = 12\, m^2$, $mC_p = 42186\, J\hspace{-0.5mm}/ \hspace{-0.5mm} K$, $U_{g1} = 6.27\, W \hspace{-0.5mm}/ \hspace{-0.5mm} K$, $U_{12} = 5.08 \, W \hspace{-0.5mm}/ \hspace{-0.5mm} K$, $U_{23} = 5.41\, W\hspace{-0.5mm}/ \hspace{-0.5mm} K$, $U_{3g} = 6.27\, W \hspace{-0.5mm}/ \hspace{-0.5mm} K$, $Q_{hAC} = 350\, W$, $Q_{dw} = 300\, W$, $Q_{Sl} = 200\, W$, and $T_{tg} = 293\, K$.

Since $\lambda(A) = \big\{ -0.052, -0.033, -0.010\big\} \subseteq \mathbb{R}^-$, $A$ is Hurwitz. Then, according to Theorem~\ref{thm: N&S resilience}, the system is not resilient, but it might be resiliently stabilizable. For the loss of any one column $C$, $\rank(B) = 3$ and we numerically verify that $-C\mathcal{W} \subseteq \interior(B\mathcal{U})$. Then, following Lemma~\ref{lemma: dim Z = rank B}, $\dim(\mathcal{Z}) = 3$, so $\interior(\mathcal{Z}) \neq \emptyset$. According to Proposition~\ref{prop: resilient stabilizability Z}, the system is resiliently stabilizable.

\begin{figure}[htbp!]
    \centering
    \includegraphics[scale=0.5]{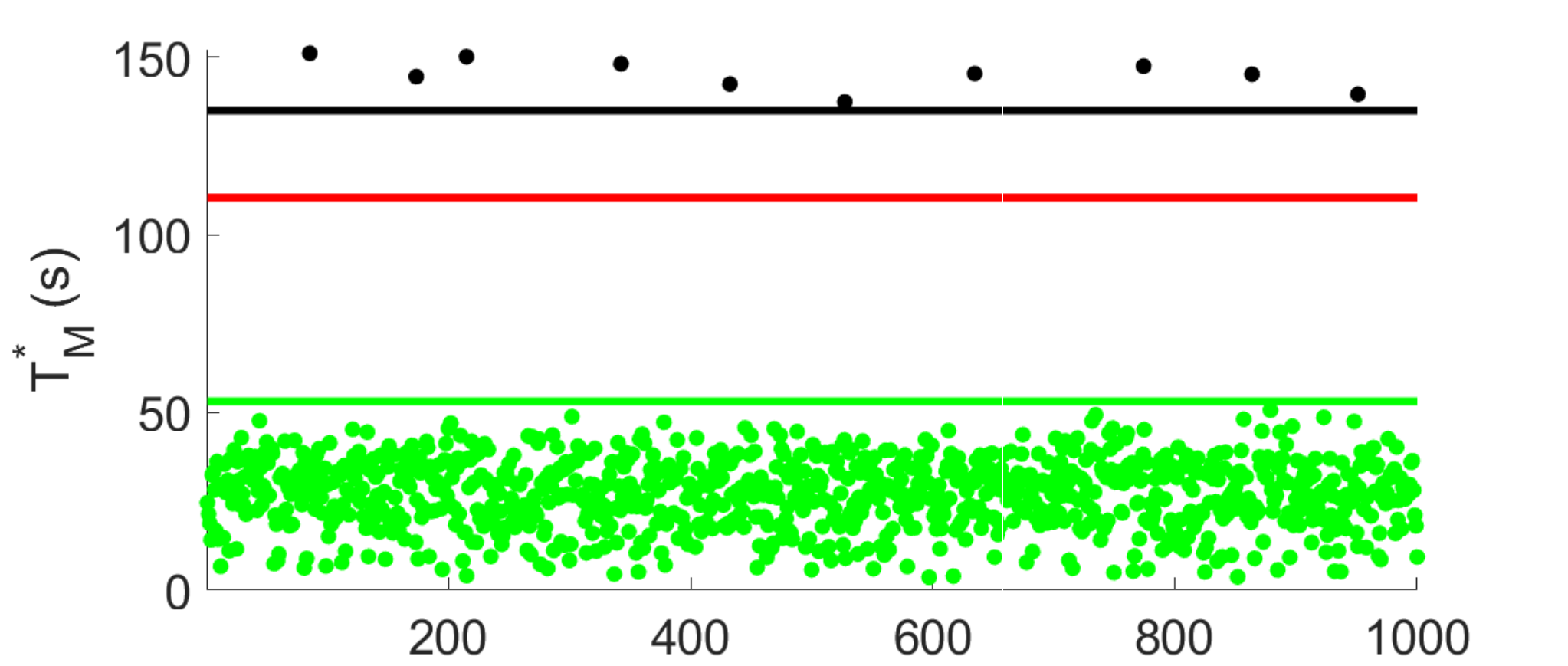}
    \caption{Bounds on the malfunctioning reach time $T_M^*(x_0)$ in red. The dots are the upper \eqref{eq: T_M ub} and lower bounds \eqref{eq: T_M lb} for 1000 stochastic pairs $(P,Q)$.
    The tightest bounds in green and black result from the ellipsoidal approximations of $\mathcal{Z}$.}
    \label{fig:T_M}
\end{figure}

The controller wants to cool the building overnight from an initial state chosen to be $x_0^\top = \big( 0.8^\circ C,\ 0.7^\circ C,\ 0.9^\circ C \big)$. However, a worker is overriding $u_{dw}^1$ by manually opening the door and window in room 1. We now compare the analytical bounds on the nominal and malfunctioning reach times of Section~\ref{sec:quantitative} with the numerical results of \cite{Eaton, Sakawa}. Our bounds require pairs $P \succ 0$ and $Q \succ 0$ solutions of $A^\top P + PA = -Q$. We generate randomly a thousand of such pairs $(P,Q)$ and compute bounds on $T_N^*$ with \eqref{eq: T_N lb} and \eqref{eq: T_N ub}, and on $T_M^*$ with \eqref{eq: T_M lb} and \eqref{eq: T_M ub}. Another way of choosing $P$ relies on the linearization of \eqref{eq: T_M lb}, which yields $T_M^* \geq \frac{\|x_0\|_P}{z_{max}^P}$. This bound is maximized when $P \succ 0$ is the tightest ellipsoidal approximation of $\mathcal{Z}$, which results in much tighter bound than stochastic $P$, as shown on Fig.~\ref{fig:T_M}.

For the given $x_0$ the best bounds on the reach times are $35.5\,s \leq T_N^*(x_0) = 42.5\,s \leq 54.1\,s$ and $53\,s \leq T_M^*(x_0) = 110.5\,s \leq 135\,s$. Then, the rooms can take up to $T_M^*(x_0)/T_N^*(x_0) = 2.6$ times longer to all reach $T_{tg}$ from the initial state $T_{tg} + x_0$ after the loss of control authority over $u_{dw}^1$, while our bounds predict a worst-case factor of $3.8$.


We were able to compute numerically $T_N^*(x_0)$ \cite{Eaton} and $T_M^*(x_0)$ \cite{Sakawa}, but accessing $r_q$ can only be done analytically with Theorems~\ref{thm: lb rq} and \ref{thm: ub rq}. Over all $x_0 \in \mathbb{R}^3$, they predict $r_q \in [0.166, 0.979]$. Hence, the loss of control over $u_{dw}^1$ can render the damaged system up to $1/0.166 = 6$ times slower to reach the target temperature from any initial state. This information could not be obtained with prior work and is the motivation for our analytical bounds in Section~\ref{sec:quantitative}.

If instead of losing control over $u_{dw}^1$ a disgruntled worker takes over the central heating/AC unit $u_{hAC}$, the rooms can take as much as $T_M^*(x_0)/T_N^*(x_0) = 4.7$ times longer to reach $T_{tg}$ from the same initial temperature, while our bound predicts a max ratio of $9.3$. These values are larger than for the loss of $u_{dw}^1$ because $Q_{hAC} > Q_{dw}$ and the central heating/AC affects directly all 3 rooms. Additionally, Theorem~\ref{thm: lb rq} yields $r_q \in [0.1, 0.37]$, so the malfunctioning controller can take between $2.7$ and $10$ times longer than nominally to enforce the target temperature from any initial condition.

\section{Conclusion and Future Work}

This paper establishes novel necessary and sufficient conditions for the resilient stabilizability and reachability of affine targets by linear systems. Additionally, we quantified the resilience of control systems to the loss of authority over some of their actuators. 

There are several avenues of future work. Building on our resilient stabilizability conditions, we have started to work on the resilience of networks to a partial loss of control authority over actuators of a subsystem.
Another interesting problem is to ensure the safety of critical systems by preventing them from visiting dangerous locations while completing their mission even after enduring a loss of control.
Future work should also aim at extending resilience theory to nonlinear systems. The main hurdle to this last project
is to establish a new proof of H\'ajek's duality theorem. Indeed, this result is essential for resilience theory and its current proof relies on the linearity of the dynamics, hence preventing a straightforward extension to nonlinear systems.

\appendix
\section{Supporting Lemmata}\label{apx:lemmas}

In this appendix we provide supporting results concerning sets $B\mathcal{U}$, $C\mathcal{W}$, and $\mathcal{Z}$ defined in Section~\ref{sec:background}. 

\begin{lem}\label{lemma: int Z non-empty iff 0 in Z}
    The interior of $\mathcal{Z}$ is non-empty if and only if $0 \in \interior(\mathcal{Z})$.
\end{lem}
\begin{pf}
     Since $\mathcal{Z}$ is convex and symmetric, so is its interior \cite{convex_geometry}. If $\interior(\mathcal{Z}) \neq \emptyset$, there exists $z \in \interior(\mathcal{Z})$, by symmetry $-z \in \interior(\mathcal{Z})$, and $0 \in \interior(\mathcal{Z})$ by convexity. The reverse implication is trivial. $\quad \blacksquare$
\end{pf}

\begin{lem}\label{lemma: relint Z non-empty iff 0 in Z}
    The following statements are equivalent:
    \begin{enumerate*}[label=(\alph*)]
        \item $0 \hspace{-0.5mm} \in \hspace{-0.5mm} \relint(\mathcal{Z})$,
        \item $0 \hspace{-0.5mm} \in \hspace{-0.5mm} \mathcal{Z}$,
        \item $\mathcal{Z} \hspace{-0.5mm} \neq \hspace{-0.5mm} \emptyset$,
        \item $\relint(\mathcal{Z}) \hspace{-0.5mm} \neq \hspace{-0.5mm} \emptyset$.
    \end{enumerate*}
\end{lem}
\begin{pf}
    Since $\relint(\mathcal{Z}) \subseteq \mathcal{Z}$, we have $(a) \implies (b)$ and trivially, $(b) \implies (c)$. Since $\mathcal{Z}$ is a convex subset of $\mathbb{R}^n$, $(c) \implies (d)$ according to Lemma~7.33 of \cite{inf_dim_analysis}. Because $\mathcal{Z}$ is convex and symmetric, so is its relative interior according to \cite{convex_geometry}. Then, the same proof as for Lemma~\ref{lemma: int Z non-empty iff 0 in Z} yields $(d) \implies (a)$ which completes the proof.  $\quad \blacksquare$
\end{pf}

\begin{defn}
     The dimension of a compact set $\mathcal{S}$ is the dimension of the smallest affine subspace (with respect to inclusion) containing $\mathcal{S}$ \cite{inf_dim_analysis}.
\end{defn}

\begin{lem}\label{lemma: dim Z = rank B}
    The relative interior of $B\mathcal{U}$ contains $-C\mathcal{W}$ if and only if $\dim(\mathcal{Z}) = \rank(B)$.
\end{lem}
\begin{pf}
    Let $q := \dim(B\mathcal{U}) \leq n$. Since $\mathcal{U} = [-1, 1]^{m-p}$, its interior is not empty in $\mathbb{R}^{m-p}$ and thus $q = \rank(B)$. Take $q$ linearly independent vectors of $B\mathcal{U}$ denoted by  $B_q := (b_1, \hdots, b_q)$ and pick $V := (v_{q+1},\hdots,v_n) \in \mathbb{R}^{n \times (n-q)}$ such that $T_b := (B_q, V)$ is invertible. Then, $T_b$ is a transition matrix with $T_b e_i = b_i$ for $i \in [\![1,q]\!]$.

    Assume first that $-C\mathcal{W} \subseteq \relint(B\mathcal{U})$. Then, there exists $\varepsilon > 0$ such that $T_b\big(\mathbb{B}^q(0, \varepsilon) \times \{0\}^{n-q}\big) \oplus -C\mathcal{W} \subseteq B\mathcal{U}$. Informally, $-C\mathcal{W}$ remains in $B\mathcal{U}$ when it is 'extended' by $\varepsilon$ in all $q$ dimensions of $B\mathcal{U}$. Because $\mathcal{Z} = \big\{ z \in \mathbb{R}^n : \{z\} \oplus -C\mathcal{W} \subseteq B\mathcal{U} \big\}$, we have $T_b\big(\mathbb{B}^q(0, \varepsilon) \times \{0\}^{n-q}\big) \subseteq \mathcal{Z}$. Then, $q \leq \dim(\mathcal{Z})$. Since $0 \in -C\mathcal{W}$, $\mathcal{Z} \subseteq B\mathcal{U}$, and hence $\dim(\mathcal{Z}) \leq q$. Thus, $\dim(\mathcal{Z}) = q = \rank(B)$.

    On the other hand, assume that $\dim(\mathcal{Z}) = q$. Since $0 \in -C\mathcal{W}$, $\mathcal{Z} \subseteq B\mathcal{U}$. Then, $\mathcal{Z}$ being of same dimension and included in $B\mathcal{U}$ yields that $(b_1,\hdots,b_q)$ is also a basis of $\Span(\mathcal{Z}) = \Image(B)$. Hence, $T_b$ is a transition matrix from $\mathbb{R}^n$ to $\Span(\mathcal{Z})$. According to Lemma~\ref{lemma: relint Z non-empty iff 0 in Z}, $0 \in \relint(\mathcal{Z})$, i.e, there exists $\delta > 0$ such that $T_b\big(\mathbb{B}^q(0, \delta) \times \{0\}^{n-q}\big) \subseteq \mathcal{Z}$. As above, the definition of $\mathcal{Z}$ yields $T_b\big(\mathbb{B}^q(0, \delta) \times \{0\}^{n-q}\big) \oplus (-C\mathcal{W}) \subseteq B\mathcal{U}$. Because $\dim( \mathbb{B}^q(0, \varepsilon)) = q = \dim(B\mathcal{U})$, we have $-C\mathcal{W} \subseteq \relint(B\mathcal{U})$. $\quad \blacksquare$
\end{pf}

\begin{lem}\label{lemma: span Z = Im B}
     If $\dim(\mathcal{Z}) = \rank(B)$, then $\Span(\mathcal{Z}) = \Image(B) = \Image(\bar{B})$.
\end{lem}
\begin{pf}
    In the proof of Lemma~\ref{lemma: dim Z = rank B} we showed that $\Span(\mathcal{Z}) = \Image(B)$. The inclusion $-C\mathcal{W} \subseteq \relint(B\mathcal{U})$ holds according to Lemma~\ref{lemma: dim Z = rank B} and yields $\Image(C) \subseteq \Image(B)$, and since $\bar{B} = [ B\ C ]$ after adequate column permutations, we have $\Image(\bar{B}) = \Image([ B\ C]) = \Image(B)$. $\quad \blacksquare$
\end{pf}

\begin{lem}\label{lemma: Z = empty}
    Set $\mathcal{Z}$ is empty if and only if set $C\mathcal{W}$ is not entirely included in $B\mathcal{U}$, i.e., $\mathcal{Z} = \emptyset \iff C\mathcal{W} \nsubseteq B\mathcal{U}$.
\end{lem}
\begin{pf}
    If $\mathcal{Z} = \emptyset$, then by definition, for all $z \in B\mathcal{U}$, there exists $w \in \mathcal{W}$ such that $z - Cw \notin B\mathcal{U}$. Taking $z = 0$ yields $C\mathcal{W} \nsubseteq B\mathcal{U}$.
    
    On the other hand, assume that there exists $w \in \mathcal{W}$ such that $Cw \notin B\mathcal{U}$. Assume for contradiction purposes that $\mathcal{Z} \neq \emptyset$. Then, we can take $z \in \mathcal{Z}$ and $z - Cw \in B\mathcal{U}$. Since $B\mathcal{U}$ is symmetric, we thus have $-z + Cw \in B\mathcal{U}$. Because $z \in \mathcal{Z}$ and $-w \in \mathcal{W}$, we also have $z + Cw \in B\mathcal{U}$. The convexity of $B\mathcal{U}$ yields $\frac{1}{2}(-z+Cw) + \frac{1}{2}(z + Cw) \in B\mathcal{U}$, i.e., $Cw \in B\mathcal{U}$ which contradicts our first assumption. Hence, $\mathcal{Z} = \emptyset$. $\quad \blacksquare$
\end{pf}

\begin{lem}\label{lemma: not stabilizable}
    If $C\mathcal{W} \nsubseteq B\mathcal{U}$, then system \eqref{eq:splitted ODE} is not resiliently stabilizable.
\end{lem}
\begin{pf}
    Since $C\mathcal{W} \nsubseteq B\mathcal{U}$, there exists $w \in \mathcal{W}$ such that $Cw \notin B\mathcal{U}$. The sets $\{Cw\}$ and $B\mathcal{U}$ are nonempty, disjoint, convex, and compact, hence they are strongly separated according to Theorem~5.79 of \cite{inf_dim_analysis}. Then, there exists $v \in \mathbb{R}^n$, $v \neq 0$, $c > 0$, and $\varepsilon > 0$ such that $\langle Cw, v\rangle \geq c + \varepsilon$, and for all $u \in \mathcal{U}$, $\langle Bu, v \rangle \leq c - \varepsilon$. Because $B\mathcal{U}$ and $C\mathcal{W}$ are symmetric, $\{-Cw\}$ and $B\mathcal{U}$ are also strongly separated by the symmetric hyperplane: $\langle -Cw, v\rangle \leq -c - \varepsilon$ and for all $u \in \mathcal{U}$, $\langle Bu, v \rangle \geq -c + \varepsilon$.
    
    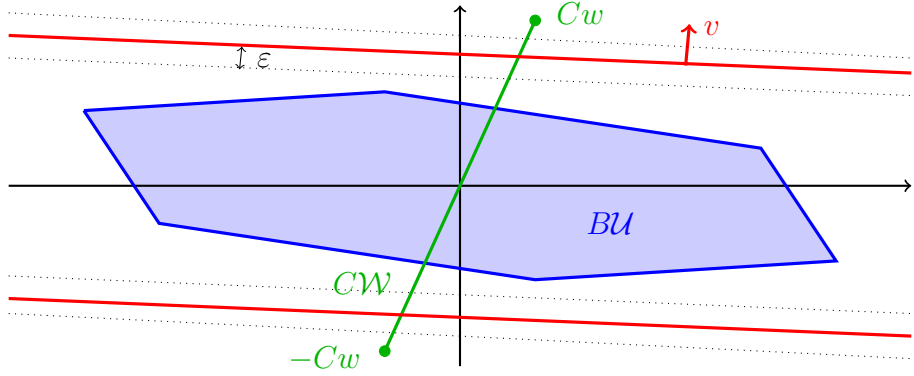
\begin{figure}[htbp!]
        \centering
        \begin{tikzpicture}[scale = 1]
            
            \filldraw[blue!20!white] (-5, 1) -- (-1, 1.25) -- (4, 0.5) -- (5, -1) -- (1, -1.25) -- (-4, -0.5) -- (-5, 1);
            
            \draw[thick, ->] (-6, 0) -- (6, 0);
            \draw[thick, ->] (0, -2.4) -- (0, 2.4);
            
            \draw[very thick, blue] (-5, 1) -- (-1, 1.25) -- (4, 0.5) -- (5, -1) -- (1, -1.25) -- (-4, -0.5) -- (-5, 1);
            \node at (2, -0.5) {\textcolor{blue}{$B\mathcal{U}$}};
            
            \draw[very thick, green!70!black] (-1, -2.2) -- (1, 2.2);
            \filldraw[green!70!black] (1, 2.2) circle (2pt);
            \filldraw[green!70!black] (-1, -2.2) circle (2pt);
            \node at (1.6, 2.3) {\textcolor{green!70!black}{$Cw$}};
            \node at (-1.8, -2.3) {\textcolor{green!70!black}{$-Cw$}};
            \node at (-1.3, -1.3) {\textcolor{green!70!black}{$C\mathcal{W}$}};
            
            \draw[very thick, red] (-6, 2) -- (6, 1.5);
            \draw[very thick, red] (-6, -1.5) -- (6, -2);
            
            \draw[very thick, red, ->] (3, 1.6) -- (3.05, 2.15);
            \node at (3.35, 2.1) {\textcolor{red}{$v$}};
            
            \draw[dotted] (-6, 2.3) -- (6, 1.7);
            \draw[dotted] (-6, 1.7) -- (6, 1.2);
            \draw[<->] (-2.93, 1.55) -- (-2.9, 1.85);
            \node at (-2.6, 1.67) {$\varepsilon$};
            
            \draw[dotted] (-6, -1.2) -- (6, -1.7);
            \draw[dotted] (-6, -1.7) -- (6, -2.3);
        \end{tikzpicture}
        \caption{Illustration of the strong separation of sets $B\mathcal{U}$ (blue) and $\{\pm Cw\}$ (green) by symmetric hyperplanes.}
        \label{fig:hyperplane}
    \end{figure}

    If $A \neq 0$, then $\|A\| > 0$. Since $v \neq 0$, we can define $r := \frac{\varepsilon}{\|v\| \, \|A\|} > 0$. We will show that if $x \in \mathbb{B}^n(0, r)$, then no controls $u \in \mathcal{U}$ can bring the state $x$ closer to the origin. Let $x \in \mathbb{B}^n(0, r)$ and first assume that $\langle x, v\rangle \geq 0$. Then, we apply the undesirable input $w$ and any control $u \in \mathcal{U}$ to system \eqref{eq:splitted ODE}
    \begin{equation*}
        \langle \dot x, v\rangle = \langle Ax, v\rangle + \langle Bu, v \rangle + \langle Cw, v \rangle \geq -\|Ax\| \, \|v\| -c + \varepsilon + c + \varepsilon \geq - \|A\| \, \|x\| \, \|v\| + 2\varepsilon \geq \varepsilon,
    \end{equation*}
    where we used the Cauchy-Schwarz inequality \cite{matrix_computations}, the definition of $\|A\|$ and $\|x\| \leq r$. Similarly, if $\langle x, v \rangle < 0$, we apply the undesirable input $-w$ and any control $u \in \mathcal{U}$ to system \eqref{eq:splitted ODE}
    \begin{equation*}
        \langle \dot x, v\rangle = \langle Ax, v\rangle + \langle Bu, v \rangle + \langle -Cw, v \rangle \leq \|A\| \, \|x\| \, \|v\| + c - \varepsilon - c - \varepsilon \leq r \|A\| \, \|v\| -2\varepsilon = -\varepsilon.
    \end{equation*}
    Thus, the state $x \in \mathbb{B}^n(0, r)$ can be pushed away from the origin along $v$. Hence, system \eqref{eq:splitted ODE} is not stabilizable.
    
    If $A = 0$, we can take any $x \in \mathbb{R}^n$ such that $\langle x, v \rangle \geq 0$ (resp. $\leq 0$) and obtain $\langle \dot x, v \rangle \geq 2\varepsilon$ (resp. $\leq -2\varepsilon$) so the same conclusion holds. $\quad \blacksquare$
\end{pf}

\section*{Acknowledgment}

This work was supported by an Early Stage Innovations grant from NASA’s Space Technology Research Grants Program, grant no. 80NSSC19K0209. This material is partially based upon work supported by the United States Air Force AFRL/SBRK under contract no. FA864921P0123.

The authors thank Dr. Bordignon and Dr. Durham for providing us with the ADMIRE model and Dr. Althoff for his help concerning zonotopes.

\begin{spacing}{0.9}
\bibliographystyle{abbrv}
\bibliography{references}

\begin{thebibliography}{10}

\bibitem{inf_dim_analysis}
C.~Aliprantis and K.~Border.
\newblock {\em Infinite Dimensional Analysis: A Hitchhiker's Guide}.
\newblock Springer, New York, 2006.

\bibitem{Zonotopes}
M.~Althoff.
\newblock On computing the {M}inkowski difference of zonotopes.
\newblock {\em arXiv preprint arXiv:1512.02794}, 2015.

\bibitem{CORA}
M.~Althoff, N.~Kochdumper, and M.~Wetzlinger.
\newblock {CORA} 2020 manual.
\newblock {\em TU Munich}, 2016.

\bibitem{Fault_Tolerant_Review}
A.~A. Amin and K.~M. Hasan.
\newblock A review of fault tolerant control systems: advancements and
  applications.
\newblock {\em Measurement}, 143:58 -- 68, 2019.

\bibitem{adaptive_control}
B.~Anderson and A.~Dehghani.
\newblock Challenges of adaptive control–past, permanent and future.
\newblock {\em Annual Reviews in Control}, 32:123 --– 135, 2008.

\bibitem{Athans}
M.~Athans.
\newblock The status of optimal control theory and applications for
  deterministic systems.
\newblock {\em IEEE Transactions on Automatic Control}, 11(3):580 -- 596, 1966.

\bibitem{ISS_thruster}
M.~Bartels.
\newblock Russia says 'software failure' caused thruster misfire at space
  station.
\newblock {\em
  https://www.space.com/space-station-nauka-arrival-thruster-fire-update},
  2021.

\bibitem{Borgest}
W.~Borgest and P.~Varaiya.
\newblock Target function approach to linear pursuit problems.
\newblock {\em IEEE Transactions on Automatic Control}, 16(5):449 -- 459, 1971.

\bibitem{IFAC}
J.-B. Bouvier and M.~Ornik.
\newblock Resilient reachability for linear systems.
\newblock In {\em 21st IFAC World Congress}, pages 4409 -- 4414, 2020.

\bibitem{TAC}
J.-B. Bouvier and M.~Ornik.
\newblock Designing resilient linear systems.
\newblock {\em IEEE Transactions on Automatic Control}, 67(9):4832 -- 4837,
  2022.

\bibitem{Maximax_Minimax_JOTA}
J.-B. Bouvier and M.~Ornik.
\newblock The maximax minimax quotient theorem.
\newblock {\em Journal of Optimization Theory and Applications}, 192:1084 --
  1101, 2022.

\bibitem{ECC}
J.-B. Bouvier and M.~Ornik.
\newblock Quantitative resilience of linear systems.
\newblock In {\em 20th European Control Conference}, pages 485 -- 490, 2022.

\bibitem{SIAM_CT}
J.-B. Bouvier, K.~Xu, and M.~Ornik.
\newblock Quantitative resilience of linear driftless systems.
\newblock In {\em SIAM Conference on Control and its Applications}, pages 32 --
  39, 2021.

\bibitem{Quantitative_Resilience}
J.-B. Bouvier, K.~Xu, and M.~Ornik.
\newblock Quantitative resilience of generalized integrators.
\newblock {\em {in review}}, https://arxiv.org/abs/2111.04163.

\bibitem{Brammer}
R.~F. Brammer.
\newblock Controllability in linear autonomous systems with positive
  controllers.
\newblock {\em SIAM Journal on Control}, 10(2):339 -- 353, 1972.

\bibitem{actuators_measures}
J.~Davidson, F.~Lallman, and T.~Bundick.
\newblock Real-time adaptive control allocation applied to a high performance
  aircraft.
\newblock In {\em 5th SIAM Conference on Control and Its Applications}, 2001.

\bibitem{Eaton}
J.~H. Eaton.
\newblock An iterative solution to time-optimal control.
\newblock {\em Journal of Mathematical Analysis and Applications}, 5(2):329 --
  344, 1962.

\bibitem{FOI_Admire}
L.~Forssell and U.~Nilsson.
\newblock {ADMIRE}: The aero-data model in a research environment version 4.0,
  model description.
\newblock Technical report, FOI - Swedish Defence Research Agency, December
  2005.

\bibitem{Girard}
A.~Girard, C.~{Le Guernic}, and O.~Maler.
\newblock Efficient computation of reachable sets of linear time-invariant
  systems with inputs.
\newblock In {\em International Workshop on Hybrid Systems: Computation and
  Control}, pages 257 -- 271. Springer, 2006.

\bibitem{matrix_computations}
G.~Golub and C.~{Van Loan}.
\newblock {\em Matrix Computations}.
\newblock John Hopkins University Press, 2013.

\bibitem{Hajek}
O.~H{\'a}jek.
\newblock Duality for differential games and optimal control.
\newblock {\em Mathematical Systems Theory}, 8(1):1 -- 7, 1974.

\bibitem{ADMIRE_2}
O.~H{\"{a}}rkeg{\aa}rd and S.~T. Glad.
\newblock Resolving actuator redundancy - optimal control vs. control
  allocation.
\newblock {\em Automatica}, 41:137 -- 144, 2005.

\bibitem{Heymann_long}
M.~Heymann, M.~Pachter, and R.~Stern.
\newblock Max-min control problems: A system theoretic approach.
\newblock {\em IEEE Transactions on Automatic Control}, 21(4):455 -- 463, 1976.

\bibitem{Isaacs_review}
Y.-C. Ho.
\newblock Review of the book {Differential Games by R. Isaacs}.
\newblock {\em IEEE Transactions on Automatic Control}, 10:501 -- 503, 1965.

\bibitem{Kalman}
R.~E. Kalman and J.~E. Bertram.
\newblock Control system analysis and design via the “second method” of
  {L}yapunov: continuous-time systems.
\newblock {\em Journal of Basic Engineering}, 82(2):371 -- 393, 1960.

\bibitem{Khalil}
H.~K. Khalil.
\newblock {\em Nonlinear Systems}.
\newblock Prentice Hall, 2002.

\bibitem{Pontryagin_difference}
I.~Kolmanovsky and E.~G. Gilbert.
\newblock Theory and computation of disturbance invariant sets for
  discrete-time linear systems.
\newblock {\em Mathematical Problems in Engineering}, 4(4):317 -- 367, 1998.

\bibitem{real_variables}
S.~G. Krantz.
\newblock {\em A handbook of real variables: with applications to differential
  equations and {Fourier} analysis}.
\newblock Springer Science \& Business Media, 2011.

\bibitem{Liberzon}
D.~Liberzon.
\newblock {\em Calculus of Variations and Optimal Control Theory: a Concise
  Introduction}.
\newblock Princeton University Press, 2011.

\bibitem{convex_geometry}
M.~Moszynska.
\newblock {\em Selected Topics in Convex Geometry}.
\newblock Springer, 2006.

\bibitem{Rechtschaffen_equivalences}
E.~Rechtschaffen.
\newblock Equivalences between differential games and optimal controls.
\newblock {\em Journal of Optimization Theory and Applications}, 18(1):73 --
  79, 1976.

\bibitem{Sakawa}
Y.~Sakawa.
\newblock Solution of linear pursuit-evasion games.
\newblock {\em SIAM Journal on Control}, 8(1):100 -- 112, 1970.

\bibitem{Schmitendorf_MaxMin}
W.~Schmitendorf and B.~Elenbogen.
\newblock Constrained max-min controllability.
\newblock {\em IEEE Transactions on Automatic Control}, 27(3):731 -- 733, 1982.

\bibitem{actuator_lock}
G.~Tao, S.~Chen, and S.~M. Joshi.
\newblock An adaptive actuator failure compensation controller using output
  feedback.
\newblock {\em IEEE Transactions on Automatic Control}, 47(3):506 -- 511, 2002.

\bibitem{temperature}
S.~H. Trapnes.
\newblock {Optimal Temperature Control of Rooms}.
\newblock Master's thesis, Norwegian University of Science and Technology,
  2012.

\bibitem{weak_robust_control}
L.~Y. Wang and J.-F. Zhang.
\newblock Fundamental limitations and differences of robust and adaptive
  control.
\newblock In {\em 2001 American Control Conference}, pages 4802 -- 4807, 2001.

\bibitem{loss_control_effectiveness}
B.~Xiao, Q.~Hu, and P.~Shi.
\newblock Attitude stabilization of spacecrafts under actuator saturation and
  partial loss of control effectiveness.
\newblock {\em IEEE Transactions on Control Systems Technology}, 21(6):2251 --
  2263, 2013.

\end{thebibliography}
\end{spacing}

\end{document}